\documentclass[fleqn,usenatbib]{mnras}
\usepackage{lmodern}
\usepackage[T1]{fontenc}
\usepackage{aecompl}

\usepackage[utf8]{inputenc}
\usepackage{bm}
\usepackage{caption}

\usepackage[dvipsnames]{xcolor}
\definecolor{mycolor_0}{RGB}{255,50,50}
\definecolor{mycolor}{RGB}{255,160,0}
\definecolor{mycolor_1}{RGB}{0,190,190}
\definecolor{mycolor_2}{RGB}{11,220,220}

\usepackage{graphicx}	
\usepackage{amsmath}	
\usepackage{amssymb}	
\usepackage{ulem}
\usepackage{rotating}
\usepackage{orcidlink}
\usepackage{xspace}

\newcommand{\tavg}{\Delta t_{\rm avg}}

\newcommand{\mol}{{\mathrm{H}_2}}

\newcommand{\ximol}{\xi_\mol}

\newcommand{\Myr}{~\text{Myr}}
\newcommand{\Gyr}{~\text{Gyr}}
\newcommand{\Msun}{~\text{M}_\odot}
\newcommand{\pc}{~\text{pc}}
\newcommand{\kpc}{~\text{kpc}}
\newcommand{\Mpc}{~\text{Mpc}}
\newcommand{\vir}{{\text{vir}}}
\newcommand{\percent}{~\text{per~cent}}

\newcommand{\firebox}{\text{FIREbox}\xspace}
\newcommand{\skirt}{\textsc{skirt}\xspace}
\newcommand{\cloudy}{\textsc{cloudy}\xspace}
\newcommand{\toddlers}{\textsc{toddlers}\xspace}
\newcommand{\AAA}{{\text{\AA}}}
\newcommand{\squiggle}{SQuIGG$\vec{L}$E\xspace}
\newcommand{\true}{`true'\xspace}

\newcommand{\fb}{{\bf{\mathrm{}}}}

\newcommand{\Fim}{{\mathcal{F}_{\,\rm Im}\,}}
\newcommand{\FimFB}{{\mathcal{F}_{\,\rm Im}^{\fb}\,}}

\title[Post-starburst galaxies in \firebox]{The Nature of Post-Starburst Galaxies: Real Deal or Masquerading Impostors?}

\author[Cenci et al.]{
\parbox{\textwidth}{
Elia Cenci$^{1,2}$\orcidlink{0000-0002-0766-1704}\thanks{E-mail:\href{mailto:elia.cenci@uzh.ch}{elia.cenci@uzh.ch}},
Robert Feldmann$^{2}$\orcidlink{0000-0002-1109-1919},
Sarah Wellons$^{3}$\orcidlink{0000-0002-3977-2724},
Jindra Gensior$^{4,2}$\orcidlink{0000-0001-6119-9883},
Luigi Bassini$^{2}$\orcidlink{0000-0002-6864-7762},
Mauro Bernardini$^{2}$\orcidlink{0000-0002-2930-9509},
Rachel Bezanson$^{5}$\orcidlink{0000-0001-5063-8254},
Jorge Moreno$^{6,7}$\orcidlink{0000-0002-3430-3232},
David J. Setton$^{8}$\orcidlink{0000-0003-4075-7393},
Lucas Tortora$^{9,2}$\orcidlink{0009-0005-8040-8325}
\vspace{6pt}
}\\
$^{1}$Department of Astronomy, University of Geneva, Chemin Pegasi 51, Versoix CH-1290, Switzerland\\
$^{2}$Department of Astrophysics, Universit\"at Z\"urich, Winterthurerstrasse 190, Zurich CH-8057, Switzerland\\
$^{3}$Department of Astronomy, Van Vleck Observatory, Wesleyan University, 96 Foss Hill Drive, Middletown, CT 06459, USA \\
$^{4}$Institute for Astronomy, University of Edinburgh, Royal Observatory, Blackford Hill, Edinburgh EH9 3HJ, UK \\
$^{5}$Department of Physics and Astronomy and PITT PACC, University of Pittsburgh, Pittsburgh, PA 15260, USA\\
$^{6}$Department of Physics and Astronomy, Pomona College, Claremont, CA 91711, USA \\
$^{7}$Carnegie Observatories, Pasadena, CA 91101, USA\\
$^{8}$Department of Astrophysical Sciences, Princeton University, 4 Ivy Lane, Princeton, NJ 08544, USA\\
$^{9}$Institute of Astronomy, University of Cambridge, Madingley Road, Cambridge CB3 0HA, UK
}

\date{Accepted XXX. Received YYY; in original form ZZZ}

\pubyear{2025}

\begin{document}
\label{firstpage}
\pagerange{\pageref{firstpage}--\pageref{lastpage}}
\maketitle


\begin{abstract} 
    Post-starburst galaxies (PSBs) are a population of galaxies with spectral and photometric features indicative of rapid quenching following a recent starburst. The origin and nature of PSBs are currently debated. For example, a number of observed PSBs unexpectedly host substantial molecular gas despite their low inferred star-formation activity. Furthermore, the relative roles of galaxy interactions and quenching mechanisms in PSBs remain unclear. We study PSBs at $z=0.7$ and $z=1$ in the \firebox cosmological simulation, selecting them primarily via their rest-frame optical photometric properties. The fraction of PSBs in \firebox broadly agrees with observations, although some candidates are clear impostors with star-formation rates comparable to star-forming galaxies of similar mass. Impostors are rich in molecular gas and have a larger near-to-mid infrared flux ratios compared to quenched PSBs in the sample. The role of galaxy interactions of PSBs in \firebox depends on their stellar mass. At low stellar masses ($\lesssim 10^{10}\Msun$), PSBs have interaction fractions comparable to those of non-PSBs in the simulation, consistent with a scenario in which stellar feedback and gas consumption drive temporary quenching of star formation. At higher stellar masses ($\gtrsim 10^{10}\Msun$), PSBs are preferentially interacting systems compared to non-PSBs, with major mergers providing the dominant contribution. We conclude that stellar feedback and galaxy interactions in \firebox can produce galaxies with observational properties akin to those of observed PSBs, many of which are actively forming stars. Additional quenching channels, such as massive black hole feedback, are likely required to explain a long-lived, quenched population of PSBs.
\end{abstract}


\section{Introduction}
Galaxies in the local Universe exhibit a strong bimodal distribution, with a cloud of blue, star-forming galaxies and a sequence of red, quiescent galaxies \citep[e.g.,][]{Blanton2003,Baldry2006,Wyder2007,Jin2014}. Besides colours and star-formation rates, this bimodality extends to a number of other galaxy properties such as their size, morphology, molecular gas content, and environment \citep[e.g.,][]{Strateva2001,Kauffmann2003a,Shen2003,Noeske2007,Wuyts2011a,Schawinski2014}. The fraction of galaxies in the red sequence is observed to increase over cosmic time, suggesting a transition between the two populations driven by the quenching of star formation \citep[][]{Bell2004,Arnouts2007,Faber2007,Jannuzi2007}

The relatively rare galaxies observed as they transition from the blue cloud to the red sequence broadly populate the so-called green valley \citep[][]{Bell2003,Wyder2007,Martin2007,Yan2009,Wong2012}. Most of the green valley galaxies are quenching by slowly exhausting of their fuel for star formation, whilst a more rare population of galaxies is undergoing a more abrupt quenching process \citep[e.g.,][]{York2000,Strauss2002,Abazajian2004,Barro2013,Barro2014,Schawinski2014,Wild2016,Blanton2017,Carnall2018,Forrest2018,Rowlands2018a,Wu2018,Belli2019,Suess2021,French2021}. The latter class of rapidly transitioning galaxies are referred to as (K+A, or E+A) post-starburst galaxies (PSBs), which are often identified based on the lack of short-lived massive stars and the presence of intermediate-age stars, indicating that star formation was quenched only recently \citep[approximately within the past $\mathrm{Gyr}$; e.g.,][]{Dressler_and_Gunn1983,Couch_and_Sharples1987,Poggianti1999,Zabludoff1996,Quintero2004,Blake2004,Tran2004,Goto2005,Goto2007b,Yang2008,Kriek2010,Young2014,Pawlik2018,Suess2021,Suess2022}. Indeed, PSBs in the local Universe have formed up to about 80 per cent of their stellar mass in one or more recent starburst events, characterised by SFRs that were a factor of about $10-100$ above the star-forming main sequence \citep[SFMS; e.g.,][]{French2018}.

Massive ($M_{\rm star}\gtrsim 10^{10}\Msun$) quiescent galaxies are already present by $z\gtrsim 4$ \citep[e.g.,][]{Straatman2014,Carnall2023,deGraaff2024}. Lower-mass ($M_{\rm star}\lesssim 10^8\Msun$) quiescent systems were also identified at similarly high redshifts \citep[][]{Strait2023,Looser2024}. The spectra of the majority of quiescent galaxies at $z\gtrsim 3$ have characteristic PSB features \citep[][]{DEugenio2020a}, and recent data from the James Webb Space Telescope \citep[JWST;][]{Gardner2006} report evidence for early quiescent galaxies at $z\gtrsim 5$ being PSBs \citep[e.g.,][]{deGraaff2024}. Furthermore, about $70\percent$ of today's ellipticals may have undergone a PSB phase by $z\sim 1$ \citep[]{Tran2004}. In general, the fraction of PSBs among the galaxy population in the field with $M_{\rm star}>10^{10}~\mathrm{M}_\odot$ increases with redshift, from $<1$ per cent at $z=0$ to about $5$ per cent at $z\sim 2$ \citep[e.g.,][]{Tran2004,Kriek2016,Wild2016,Zahid2016,Rowlands2018a,Belli2019,Suess2022,Setton2023,Park2024}. Therefore, understanding the nature of PSBs can unveil the origin of many quiescent galaxies, especially at high-redshift \citep[e.g.,][]{Wild2016}.

Given their low SFRs, quiescent galaxies and PSBs are thought to host small molecular gas fractions. Contrary to theoretical expectations, recent observations show that CO-traced molecular gas can be surprisingly abundant in selected PSBs at different redshifts \citep[e.g.,][]{Dupraz1990,
Kohno2001,Roseboom2009,French2015,Rowlands2015,Alatalo2016a,Alatalo2016b,Alatalo2016c,Suess2017,Yesuf2017,Smercina2018,Yesuf_and_Ho2020,Belli2021,Morishita2021,Williams2021,Bezanson2022,Smercina2022,Baron2023,French2023}, declining over $100-300\Myr$ time-scales \citep[e.g.,][]{French2015,Rowlands2015,Alatalo2016c,Davis2019,Li2019a,French2021,Bezanson2022}. However, observations of dense HCO-traced gas reveals that many PSBs have low or negligible HCO/CO ratios, suggesting that their gas may be in a diffuse state, and hence not involved in star formation \citep[e.g.,][]{French2018,French2023}. 

Whether PSBs are truly quenched and stopped forming stars is also unclear. Infrared and radio observations reveal significant obscured star formation in PSBs selected based on their optical spectra \citep[e.g.,][]{Smail1999,Poggianti_and_Wu2000,Smercina2018,Baron2022,Baron2023}. Therefore, depending on the specific selection criteria, a non-negligible fraction (about 4 to 30 \percent) of PSBs might host on-going star formation \citep{Baron2023}. These ‘impostor' PSBs with obscured star formation might also harbour large molecular gas reservoirs \citep[e.g.,][]{French2015,French2021,Baron2022,Baron2023}. Nonetheless, there might be a substantial fraction of PSBs that do not host any obscure star formation \citep[][]{Wild2024}.

The formation of PSBs is still unclear. A PSB phase may result from a number of different mechanisms, including ram pressure stripping by infall onto galaxy clusters \citep[e.g.,][]{Blanton_and_Moustakas2009} or intra-group interactions \citep[e.g.,][]{Zabludoff1996,Zabludoff_and_Mulchaey1998,Alatalo2015}. The presence of clear tidal features and disturbed kinematics and morphologies in isolated PSBs in the field also point towards a merger-induced origin \citep[e.g.,][]{Zabludoff1996,Yang2004,Yang2008}, especially in massive systems \citep[e.g.,][]{Ellison2024}. In this scenario, merging galaxies would undergo an intense starburst event before entering the PSB phase and then finally settling into an early-type galaxy, as suggested by numerical simulations \citep[e.g.,][]{Barnes1988,Barnes_and_Hernquist1991,Barnes_and_Hernquist1996,Mihos_and_Hernquist1996,Hopkins2006a,Di_Matteo2007,Cox2008,Genzel2010,Ellison2013,Capelo2015,Renaud2014,Moreno2015,Hopkins2018,Pan2018,Moreno2019,Renaud2019,Moreno2021,Segovia_Otero2022}. The increase with redshift of the PSB fraction is in broad agreement with the observed and simulated merger rates, supporting the merger-driven pathway to the formation of PSBs \citep[e.g.,][]{Bekki2005,Snyder2011,Davis2019}. However, theoretical and observational work show that not all galaxy mergers and interactions can trigger a starburst event and the subsequent quenching of star formation \citep[e.g.,][]{Bergvall2003c,Bekki2005,Di_Matteo2007,Di_Matteo2008,Hopkins2009a,Wild2009,Snyder2011,Sparre_and_Springel2016,Fensch2017,Wilkinson2018,Pawlik2019,Pearson2019,Diaz-Garcia_and_Knapen2020,Shah2020,Wilkinson2022,Cenci2024a,Li2023a}.


Some PSBs display luminosities and emission features typical of active galactic nuclei (AGN), that likely act to quench star formation \citep[e.g.,][]{Caldwell1996,Yan2006,Yang2006,Wild2007,Georgakakis2008,Brown2009,Lemaux2010,Kocevski2011,Mendel2013,De_Propris_and_Melnick2014,French2015,Rowlands2015,Alatalo2016a,Alatalo2016b,Maltby2019,Greene2020,French2021}. AGN feedback can effectively quench star formation by driving turbulence in the interstellar gas \citep[e.g.,][]{Alatalo2014,Smercina2018} or strong outflows \citep[e.g.,][]{Zheng2020}. However, the causal connection between PSBs and AGN is still debated \citep[e.g.,][]{Brown2009,Wild2010,Alatalo2011,Nielsen2012,De_Propris_and_Melnick2014,Alatalo2015,Alatalo2017,Meusinger2017,Davis2019,French2021}. Stellar winds are also able launch outflows with high mass-loss rates \citep[e.g.,][]{Hopkins2012b,Bolatto2013}, as recently observed in many PSBs \citep[e.g.,][]{Geach2014,Baron2017,Baron2018,Geach2018,Baron2020,Baron_and_Poznanski2017}. At high redshift ($z\gtrsim 5$), low mass ($M_{\rm star} \sim 10^7 - 10^9\Msun$) galaxies might be temporarily $\text{(mini-) quenched}$ due to their bursty star formation activity \citep[e.g.,][]{Dome2023,Baker2025}. This behaviour is also described by recent numerical work \citep[see, e.g.,][who analysed galaxies in the THESAN-ZOOM simulations.]{McClymont2025a,McClymont2025b}.

However, stellar feedback might not be energetic enough to effectively eject the molecular gas reservoir available for further star formation, especially in massive galaxies \citep[e.g.,][]{Veilleux2005}. The role of this feedback channel in the formation of PSBs is still not clear \citep[e.g.,][]{McQuinn2010,Wild2010,French2021}.

In this work, we characterise the population of PSBs in the \firebox cosmological volume simulation \citep{Feldmann2023}, selected using observational criteria that are exclusively based on their photometry. We aim to understand the mechanisms responsible for initiating the PSB phase and we provide constraints on the fraction of possibly misidentified PSBs in observations at $z=0.7$ and $z=1$. \firebox simulates galaxy formation at a high dynamic range and incorporates an accurate physical treatment of the interstellar medium (ISM) as part of the \textit{Feedback In Realistic Environments} (FIRE) project \citep{Hopkins2014,Hopkins2018}, hence being well suited to study the evolution of the structure and physical properties of galaxies in a cosmological environment and over a broad range of spatial and temporal scales. In Section~\ref{sec:methods}, we describe our modelling and sample selection. In Section~\ref{sec:results}, we show the properties of PSBs selected in \firebox and compare them with observations. In Section~\ref{sec:discussion}, we discuss the possible impostor PSBs' contribution in observations and the importance of black hole feedback in originating PSBs in the Universe.

\section{Methods}\label{sec:methods}

\subsection{\firebox}\label{sec:methods_firebox}
The \firebox simulation \citep[a cosmological $\left(22.1~\text{cMpc}\right)^3$ volume with periodic boundaries;][]{Feldmann2023} is part of the \textit{Feedback In Realistic Environments}\footnote{\url{https://FIRE.northwestern.edu}} (FIRE) project \citep{Hopkins2014,Hopkins2018}. Initial conditions at $z=120$ are created with MUlti Scale Initial Conditions \citep[MUSIC;][]{Hahn_and_Abel2011} using cosmological parameters consistent with Planck 2015 results \citep{Alves2016}: $\Omega_{\rm m}=0.3089$, $\Omega_\Lambda=1-\Omega_{\rm m}$, $\Omega_{\rm b}=0.0486$, $h=0.6774$, $\sigma_8=0.8159$, $n_{\rm s}=0.9667$ and a transfer function calculated with \textsc{camb}\footnote{\url{http://camb.info}} \citep{Lewis2000,Lewis2011}.

The simulation is run with \textsc{gizmo} \citep{Hopkins2015}. Gravitational forces between particles are computed using a modified version of the parallelisation and tree gravity solver of GADGET-3 \citep{Springel2005b}, that allows for adaptive force softening. Hydrodynamics is solved with the meshless-finite-mass method \citep{Hopkins2015}.

\firebox is run with the FIRE-2 model that includes gas cooling and heating, star formation, and stellar feedback \citep{Hopkins2018}. Feedback from supermassive black holes is not included. Gas cooling down to 10 K naturally results in a multi-phase interstellar medium with a cold component. Star formation occurs in dense ($n>300$ cm$^{-3}\equiv n_{\rm SF}$), self-shielding, self-gravitating, and Jeans-unstable gas, with an instantaneous, expected star formation rate:
\begin{equation}
   \mathrm{SFR} = \ximol \, m_{\rm gas} \; t^{-1}_{\rm ff} ~,
   \label{eqn:SFR_prescription}
\end{equation}

\noindent where $m_{\rm gas}$ is the mass of the gas particle, $\ximol$ is the fraction of molecular gas in the particle, and $t_{\rm ff}$ is the free-fall time, i.e. the time it would take for the gas particle to collapse under its self-gravity:
\begin{equation}
   t_{\rm ff} = \sqrt{\frac{3\pi}{32 G \rho}\,\,}~,
   \label{eqn:tff}
\end{equation}

where $\rho$ is the local mass density of the gas particle and $G$ is Newton's gravitational constant. Gas is converted into stars with a $100\percent$ efficiency per free-fall time. The integrated local star formation efficiency is lower due to stellar feedback, consistent with the Schmidt relation \citep{Schmidt1959,Kennicutt1998b,Orr2018}. Stellar feedback prescriptions include energy, momentum, mass, and metal injections from supernovae (type II and type Ia) and stellar winds (OB and AGB stars). Radiative feedback (including photo-ionisation and photo-electric heating) and radiation pressure from young stars is considered in the Locally Extincted Background Radiation in Optically thin Networks (LEBRON) approximation \citep{Hopkins2012a}. The FIRE-2 model has been extensively validated in a number of publications analysing properties of galaxies across a range in stellar masses and numerical resolutions \citep{Wetzel2016,Hopkins2018,Ma2018b}. Specifically, \firebox reproduces key observed galaxy properties \citep[see][for further details]{Feldmann2023}{}{} and has been further validated in several recent studies \citep[e.g.,][]{Bernardini2022,Rohr2022,Bassini2023,Gensior2023a,Bassini2024,Cenci2024a,Cenci2024b,Gensior2024,Tortora2024,Bernardini2025}{}{}.

At the initial redshift, \firebox contains $N_{\rm b}=1024^3$ gas and $N_{\rm DM}=1024^3$ dark matter particles, corresponding to a mass resolution of $m_{\rm b}=6.3\times 10^4\Msun$ for baryonic (gas and star) particles and $m_{\rm DM}=3.3\times 10^5\Msun$ for dark matter particles. The softening length is fixed to $12\pc$ (physical, up to $z=9$; comoving for $z>9$) for stars and $80\pc$ for dark matter. The force softening of gas particles is adaptive and coupled to their smoothing length down to a minimum of $1.5\pc$, reached only in the dense ISM. The force resolution is set such that the highest density we formally resolve is 1000 times the star formation threshold \citep[][]{Hopkins2018}.

\subsection{Halo and galaxy properties}\label{sec:methods_definitions}
To identify dark matter haloes in the simulation we employ the AMIGA Halo Finder (AHF)\footnote{\url{http://popia.ft.uam.es/AHF}}, using the AHF's maximum density (MAX) setting for the identification of halo centres. \citep[][]{Gill2004,Knollmann2009}. We only consider haloes containing at least 100 particles of any type, which corresponds to a minimum halo mass of $M_\vir\sim 10^7\Msun\,h^{-1}$. The halo radius $R_\vir$ is defined based on the virial overdensity criterion, so that the halo virial mass is:
\begin{equation}
   M_\vir = \frac{4\pi}{3} \Delta\left(z\right) \rho_{\rm m}\left(z\right) R_\vir^3
   ~,
   \label{eqn:Mvir}
\end{equation}
\noindent where $\rho_{\rm m}\left(z\right)$ is the critical density at a given redshift,
\begin{equation*}
    \Delta\left(z\right)=\left( 18\pi^2 - 82 \Omega_\Lambda\left(z\right) - 39\left[\Omega_\Lambda\left(z\right)\right]^2 \right)/\Omega_{\rm m}\left(z\right)
    ~,
\end{equation*}
is the overdensity parameter, and $\Omega_\Lambda\left(z\right)$, $\Omega_{\rm m}\left(z\right)$ are the cosmological density parameters at redshift $z$ \citep[][]{Bryan_and_Norman1998}.

The total radius of galaxies is defined as $10\percent$ of their halo virial radius \citep[][]{Price2017}. We can then compute half-mass radii for different components (e.g., stars, molecular gas, neutral gas) by linearly interpolating between log-binned radii and the cumulative mass. The galaxy stellar mass, $M_{\rm star}$, as well as other galaxy properties such as total gas masses and SFRs, are calculated by collecting all particles within $0.1 R_\vir$. Within any given radius, we compute SFRs averaged over a time-scale $\tavg$ as follows:
\begin{equation}
    \mathrm{SFR}_{\tavg} = M_{\rm star}\left(\rm{age}\leq\tavg\right)/\tavg
    ~,
\end{equation}
\noindent where $M_{\rm star}\left(\mathrm{age}\leq\tavg\right)$ is the sum of the at-birth masses of stars that formed within the past time interval $\tavg$. Different values for the averaging time $\tavg$ can be associated with different observational tracers for SFR \citep[e.g.,][]{Sparre2017,Flores_Velazquez2021}. We choose $\tavg=5,20,\,\mathrm{and}\,100\Myr$, approximating the characteristic time-scales probed by the most extensively employed SFR tracers, e.g., H$\alpha$ nebular emission and infrared/ultraviolet continuum from dust and young stars \citep{Calzetti2013}.

For this work, we analyse galaxies in \firebox at redshift $z=0.7$ and $z=1$ and with stellar masses $M_{\rm star}\geq 3\times 10^9\Msun$, to compare the population of PSBs in \firebox with existing recent observations \citep[see, e.g.,][]{Suess2022}, and further aim to constrain the fraction of galaxies being incorrectly selected as PSBs based on their observational properties.

\subsection{Radiative transfer}\label{sec:methods_RT}
To obtain the spectral energy distributions (SED) and intensity of the interstellar radiation field (ISRF) associated with \firebox galaxies in post-processing, we make use of the 3D Monte Carlo dust radiative transfer code \skirt\footnote{\url{https://github.com/SKIRT/SKIRT9}} \citep[version 9;][]{Camps_and_Baes2020}, that includes a full treatment of dust absorption, scattering, and re-emission. 
Radiative transfer is performed on a given wavelength grid. For each wavelength, a number of photon packets are launched during the stellar emission and dust re-emission steps of the radiative transfer simulation. In order to produce SEDs and images, a detector is virtually placed at an arbitrary distance that collects the light along a given line of sight.

For this work, radiative transfer is performed on a wavelength grid consisting of a total of 1048 wavelengths. We first consider a coarse grid of 250 wavelengths between 500 and $10^7$ $\AAA$. On top of that, we use an additional fine grid of 400 equally spaced (in logarithmic space) wavelengths in a 80 $\AAA$ window around the rest-frame H$\alpha\,6565$ and H$\delta\,4102$ lines, which are especially important for the selection of PSBs. The wavelengths in the coarse grid that would fall in either ranges of this more fine-spaced grid are not considered, reducing the total number of wavelengths from 1050 to 1048. For each wavelength in the grid, we launch $10^8$ photon packets during the stellar emission and dust re-emission steps of the radiative transfer simulation.

Additionally, for every galaxy we also run \skirt to retrieve information on their infra-red spectrum (e.g., dust emission at $24~\mu\mathrm{m}$). Specifically, for these additional runs, we use a grid of 3000 linearly spaced wavelengths between $1~\mu\mathrm{m}$ and $30~\mu\mathrm{m}$  (resolution of 100 $\AAA$), and launching $10^7$ virtual photon packets.

For the stellar emission, we employ two different approaches based on the age of stars in the simulations. For young stars with ages $\leq 20\Myr$, we use the \toddlers \citep[][]{Kapoor2023} SED family within \skirt. The \toddlers emission library provides SEDs for individual spherical star-forming clouds, explicitly tracking the time evolution of their physical properties, such as mass, number density, star formation efficiency, and metallicity. The evolving cloud structure in \toddlers is passed to the ab initio spectral synthesis and photo-ionization code \cloudy\footnote{\url{https://gitlab.nublado.org/cloudy/cloudy}} \citep[][]{Ferland2017}, in order to retrieve emission lines' properties. The library \skirt accepts a set of parameters controlling the cloud properties at the onset of star formation. We assume a maximum star-formation efficiency $\rm{SFE}=0.15$, a cloud number density equal to the threshold density for star formation in the simulation ($n_{\rm SF}=300~\mathrm{cm}^{-3}$) and a cloud mass equal to 100 times the current mass of the source star particle. These parameter choices effectively increase the stellar mass contributing to the ionising flux and were calibrated so that the intrinsic SFR in \firebox (averaged over the past $20\Myr$) approximately matches the SFR inferred from the $\rm H\alpha$ luminosity using $\mathrm{SFR}=8\times 10^{-42}\left(L_{\rm{H}\alpha}/\mathrm{erg\,s^{-1}}\right)~\Msun\,\mathrm{yr}^{-1}$ \citep[][]{Kennicutt1998a}. The use of \toddlers enables a more physical representation of young stellar populations and includes the contribution from nebular emission lines from $\mathrm{HII}$ regions. For older stars, with ages $>20\Myr$, we use the stellar evolution and spectral synthesis models from within the BPASS \citep[v2.2.1;][]{Eldridge_and_Stanway2009,Eldridge_and_Stanway2012,Stanway_and_Eldridge2016,Eldridge_and_Stanway2017,Stanway_and_Eldridge2018} SED family, to compute the stellar luminosities, given their masses, ages and metallicities.

In order to produce SEDs and images, we place a mock detector at an arbitrary distance of $10\Mpc$ that collects the light along a given line of sight. For each galaxy we considered a total of 14 viewing angles in order to study the differences in the observed spectral features and colours (see Section~\ref{sec:methods_colours}) that can arise from the specific geometry of the dust content. Of these viewing angles, 12 are uniformly distributed on the 3D sphere while the additional north and south polar directions added to explicitly sample face-on orientations.

\subsection{Colours}\label{sec:methods_colours}

\begin{figure*}
    \centering
    \includegraphics[width=.495\hsize]{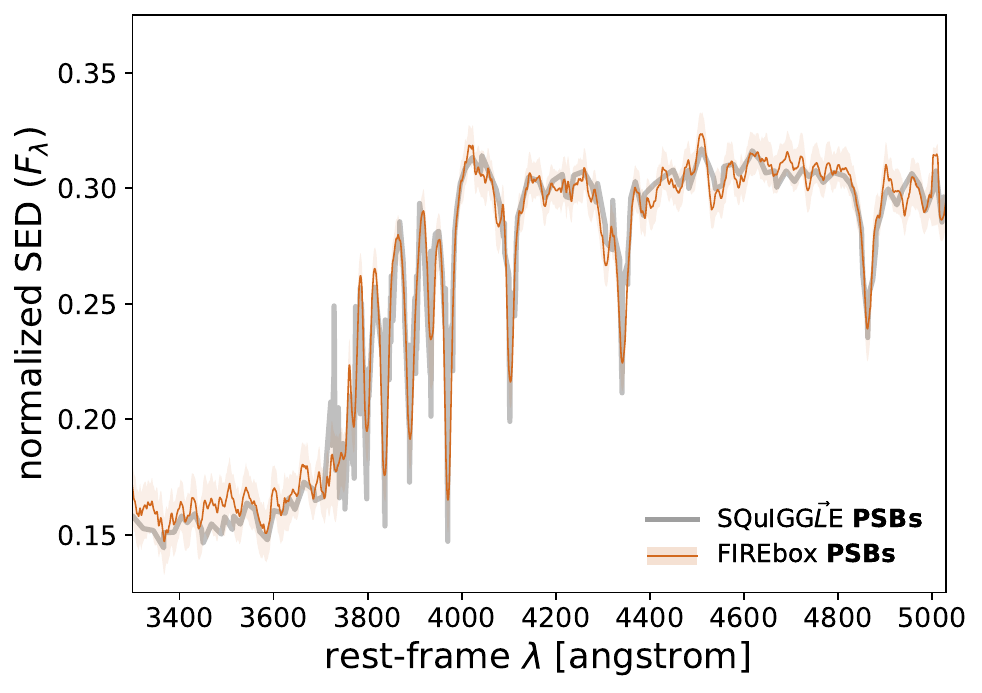}
    \includegraphics[width=.495\hsize]{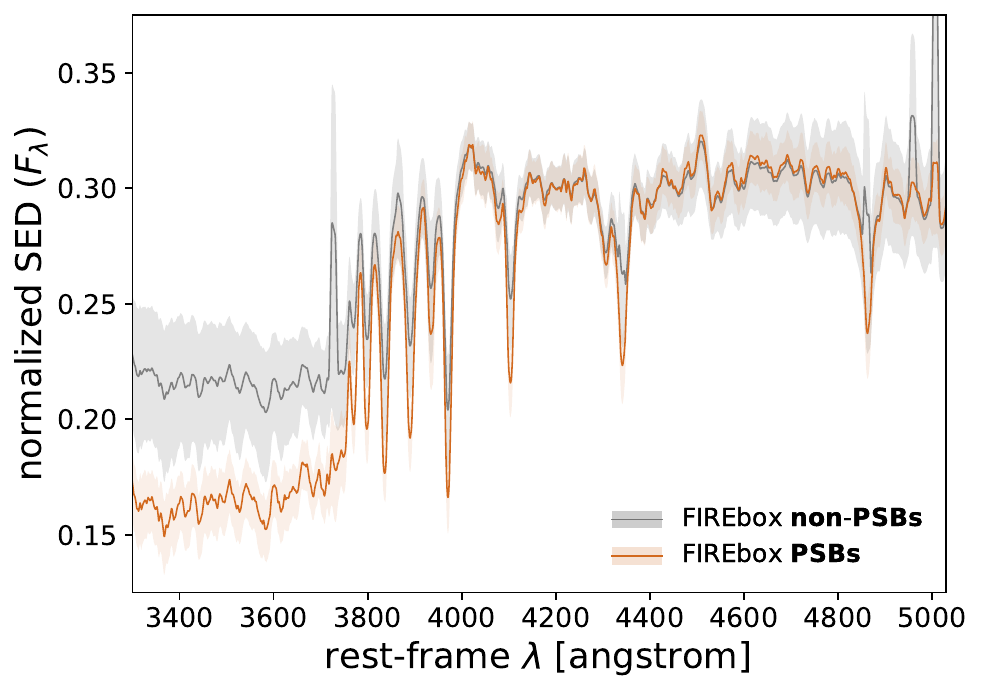}
    \caption{In the left panel, we compare the average stacked spectral energy distributions (SEDs; re-normalized) of PSBs from the \squiggle sample ($0.5<z<1$) of \citet[][]{Suess2022} (grey line) and PSBs selected in \firebox based on their photometric properties (orange line; $z=0.7$ and $z=1$, \citet{Kriek2010} selection criterion). The shaded area represent the 16th to 84th percentiles of the sample. All SEDs are normalised by the average value at rest-frame $\lambda=4200-4250\AAA$. In the right panel, we compare \firebox PSBs and a sample of redshift- and mass-matched non-PSBs (grey line). Overall, the SEDs of \firebox PSBs and \squiggle PSBs are in good agreement. On average, non-PSBs have refilled Balmer lines and less pronounced drops blue-wise of the Balmer break.}
    \label{fig:PSBs_SED}
\end{figure*}

From the SEDs generated by \skirt, we can measure the colours of galaxies in \firebox. The total flux received in a given spectral band $\mathcal{B}$ is given by:
\begin{equation}
    F_{\mathcal{B}}
    = \frac{\int_0^\infty\frac{\mathrm{d}\nu}{\nu}\,F_\nu\,\tilde{n}_\mathcal{B}\left(\nu\right)}{\int_0^\infty\frac{\mathrm{d}\nu}{\nu}\,\tilde{n}_\mathcal{B}\left(\nu\right)}
    \propto \frac{\int_0^\infty\mathrm{d}\lambda\,\lambda\,F_\lambda\,n_\mathcal{B}\left(\lambda\right)}{\int_0^\infty\frac{\mathrm{d}\lambda}{\lambda}\,n_\mathcal{B}\left(\lambda\right)}
    ~,
\end{equation}
\noindent where $F_\nu$ and $F_\lambda$ are the spectral flux density per unit frequency $\nu$ and wavelength $\lambda$, respectively. The functions $\tilde{n}_\mathcal{B}\left(\nu\right)$ and $n_\mathcal{B}\left(\lambda\right)$ are the equal-energy filter response functions (in electrons per incident photon) characterizing the $\mathcal{B}$ band in $\nu$ and $\lambda$ space, respectively. A colour is given by the ratio of the total fluxes received in two different bands, say $\mathcal{B}_1$ and $\mathcal{B}_2$:
\begin{equation}
    \mathcal{B}_1-\mathcal{B}_2 \equiv -2.5\log_{10}\left(F_{\mathcal{B}_1}/F_{\mathcal{B}_2}\right)
    ~.
\end{equation}

\subsection{Galaxy interactions}\label{sec:methods_interactions}
Central galaxies are classified as interacting if, at any point in time in a given period of time $\Delta t_{\rm int}$ in their past, there exists at least one other galaxy with a given stellar mass ratio and centre-to-centre distance. The stellar mass ratio $q_{\rm star}$ of the candidate satellite galaxy and the considered central galaxy defines whether the latter is experiencing a major ($q_{\rm star}>1:4$) or minor ($q_{\rm star}>1:10$) interaction \citep[e.g.,][]{Cox2008,Jackson2019}, while their relative distance $D$ between their centres defines the kind of interaction: merger ($D<20\kpc$), close passage ($20\kpc<D<50\kpc$), or fly-by ($50\kpc<D<100\kpc$) \citep[e.g.,][]{Benavides2021,Haggar2021}.

\subsection{Most recent starburst}\label{sec:methods_burst}

The burst age of our \firebox galaxies is an estimate for when they experienced their most recent major starburst event. For any given galaxy, we calculate its luminosity-weighted average stellar age $t_{L}$ and define the burst age as the time of maximum SFR (averaged over $\tavg = 20\Myr$), in a time period of $150\Myr$ around (centred on) $t_{L}$. This time window is chosen to be approximately twice the median time-scale for the SFR decline after the peak of a strong starburst event (about $70\Myr$), which is generally larger than the time between the peak of two subsequent starburst events \citep[][]{Cenci2024a}. 

We use the stellar evolution and spectral synthesis models from Binary Population and Spectral Synthesis\footnote{\url{https://bpass.auckland.ac.nz/9.html}} (BPASS; version v2.2.1) \citep{Eldridge_and_Stanway2009,Eldridge_and_Stanway2012,Stanway_and_Eldridge2016,Eldridge_and_Stanway2017,Stanway_and_Eldridge2018} to compute the luminosity associated with all star particles (representing entire stellar populations) in our galaxies, given their ages and metallicities. Luminosities for each population are calculated by integrating the modelled SEDs provided by BPASS over the wavelength range of $1-10^5\AAA$. We use BPASS models that assume a \citet{Chabrier2003} initial mass function (IMF) and checked that using different IMFs \citep[e.g.,][]{Salpeter1955} does not qualitatively affect our specific analysis, results and conclusions. 

We define the burst mass fraction of our galaxies as the fraction of their current stellar mass formed during the $150\Myr$ around their most recent starburst. In other words, a burst mass fraction is computed as the ratio between the total mass of stars with ages within $75\Myr$ from the galaxy burst age, divided by the current total stellar mass.

To enable a more direct comparison with observations, we also calculate the look-back time $t_{\rm PSB,90}$ when our galaxies assembled  $90\percent$ of the stellar mass that they have formed in the past $\Gyr$. To compute $t_{\rm PSB,90}$, we follow an archaeological approach, using the ages and birth masses of all stars present in the galaxy at the time when it is observed, i.e. at either $z=0.7$ or $z=1$. Our main conclusions are robust to increasing the threshold for the recently formed stellar mass fraction beyond $90\percent$. However, by doing so we are increasingly biased toward short timescales since residual star formation can persist even after a sharp decline in the SFR of our galaxies. In contrast, observational estimates of star-formation histories derived from SED fitting may tend to disfavour such short timescales \citep[][]{Setton2025}.

\subsection{Post-starburst selection}\label{sec:methods_samples}

\begin{figure}
    \centering
    \includegraphics[width=\hsize]{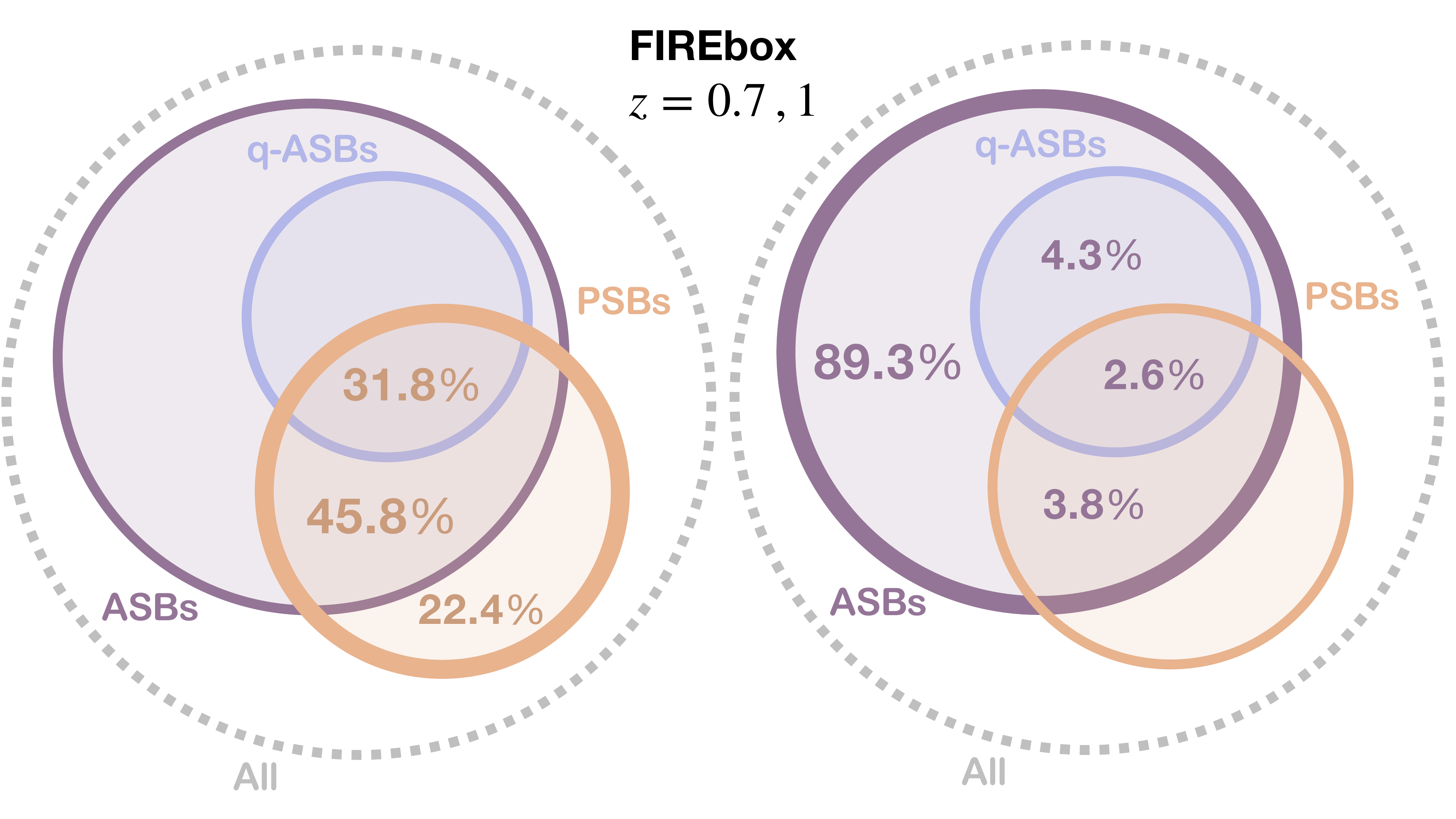}
    \caption{Schematic summary of the fraction of $z=0.7,1$ galaxies in \firebox with stellar mass $M_{\rm star}>3\times 10^{9}\Msun$, that are selected as PSBs, after-starburst galaxies (ASBs), and (temporarily) quenched ASBs (\text{q-ASBs}). \textit{Left}: Fraction of ASBs and \text{q-ASBs}, given that they are selected as PSBs in \firebox. Among PSBs, about $77.6\percent$ are ASBs and only about $31.8\percent$ are also \text{q-ASBs}. \textit{Right}: Fraction of PSBs and \text{q-ASBs}, given that they are selected as PSBs in \firebox. Among ASBs, about $6.4\percent$ are PSBs and about $6.9\percent$ are \text{q-ASBs}. Furthermore, only about $38.1\percent$ of \text{q-ASBs} are also selected as PSBs.}
    \label{fig:PSBs_impostor_frac}
\end{figure}

\renewcommand{\arraystretch}{1.5} 
\begin{table*}
    \centering
    \caption{Summary of the relevant classes of galaxies used in this work, with their fractional contribution to either the entire galaxy population or the population of PSBs selected in \firebox.}
    \label{tab:defs}
    \begin{tabular}{|p{1.5cm}|p{6.5cm}|ccccc}
        \hline
        Class, \textbf{X} & Definition
        && $\mathcal{P}\left(\,\text{\textbf{X}}\,\right)$
        & $\mathcal{P}\left(\,\text{\textbf{X}}\lvert\text{PSB}\,\right)$
        & $\mathcal{P}\left(\,\text{\textbf{X}}\lvert\neg\text{PSB}\,\right)$
        & $\mathcal{P}\left(\,\text{\textbf{X}}\lvert\text{SF}\,\right)$ \\
        \hline

        \textbf{All}
        & All galaxies considered in this study: $z=0.7,1$ galaxies in \firebox with stellar mass $M_{\rm star}>3\times10^9\Msun$. We have a total of 310 galaxies (166 at $z=0.7$ and 144 at $z=1$), each with 14 associated SEDs from different viewing angles.
        && $1$ & $1$ & $1$ & $1$ \\

        \textbf{SF}
        & Star-forming galaxies, with a current $\mathrm{sSFR}\geq3\times 10^{-11}~\mathrm{yr}^{-1}$ (calculated with a SFR averaging time of $\tavg=20\Myr$; see Section~\ref{sec:methods_ASBs}).
        && $0.971$ & $0.681$ & $0.980$ & $1$ \\
        
        \textbf{PSB}
        & Post-starburst galaxies selected with the \citet{Kriek2010} criterion and assuming an aperture of 7 kpc (in the fiducial sample). Different projections (14 available) of the same galaxy (i.e. different viewing angles) are counted separately, as individual galaxies.
        &&  $0.031$ & $1$ & $0$ & $0.022$ \\
        
        \textbf{ASB}
        & After-starburst galaxies: have a current SFR at least four times lower than the peak SFR during the most recent starburst event (see Section~\ref{sec:methods_ASBs}).
        &&  $0.374$ & $0.776$ & $0.361$ & $0.359$ \\

        \textbf{q-ASB}
        & Quenched after-starburst galaxies: ASBs that also have a present $\mathrm{sSFR}<3\times 10^{-11}~\mathrm{yr}^{-1}$: \[\text{q-ASB}=\text{ASB}\,\cap\,\neg\text{SF}~.\]
        &&  $0.026$ & $0.318$ & $0.016$ & $0$ \\

        \textbf{Im}
        & Impostor post-starburst galaxies: PSBs that have photometric properties of PSBs but are currently star-forming or did not experience a previous burst phase. In this work, impostor post-starburst galaxies are defined as PSBs that are not \text{q-ASBs}: \[\text{Im}=\text{PSB}\,\cap\,\neg\text{q-ASB}~.\]
        && $0.021$ & $0.682$ & $0$ & $0.022$ \\

        \hline
    \end{tabular}    
\end{table*}

\begin{figure}
    \centering
    \includegraphics[width=\hsize]{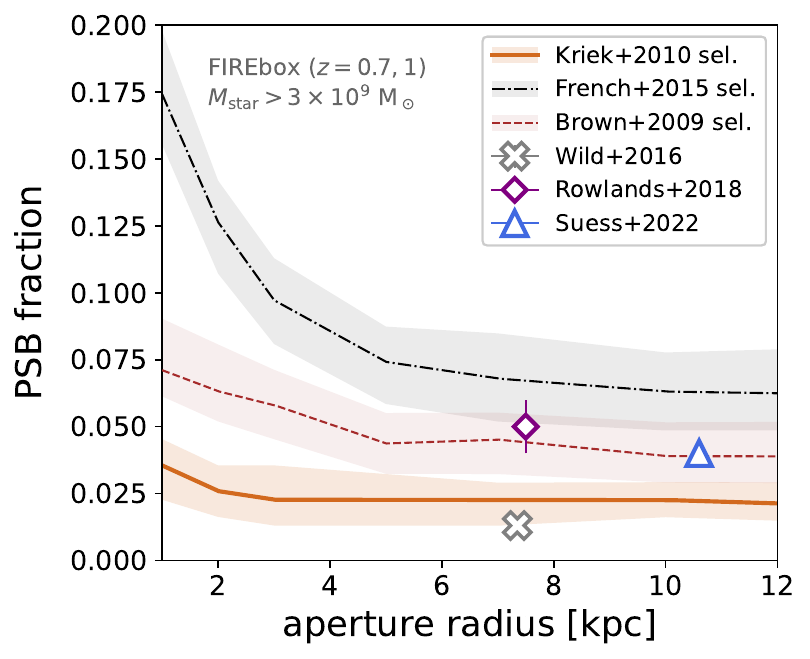}
    \caption{Median PSB fraction among all \firebox galaxies at $z=0.7,1$, as a function of the aperture radius (in physical kpc) used to produce their mock SEDs and colours. Different lines refer to different PSB selection criteria: $\left(U-B\right)_{\rm rest}$ and $\left(B-V\right)_{\rm rest}$ colours-based \citep[orange, solid line;][]{Kriek2010}{}{}; $\mathrm{EW}\left(\mathrm{H}\alpha\right)<3$ (in emission) and $\mathrm{Lick}~\mathrm{H}\delta_{\rm A}>4$ \citep[black, dash-dotted line;][]{French2015}{}{}; $\log\left[\mathrm{EW}\left(\mathrm{H}\alpha\right)\right] < 0.2\times \mathrm{Lick}~\mathrm{H}\delta_{\rm A}$ and $\mathrm{EW}\left(\mathrm{H}\delta_{\rm A}\right)>3$ \citep[red, dashed line;][]{Brown2009}{}{}. The shaded areas represent the bootstrapped variation of the median PSB fraction, between the 16th to 84th percentiles. We compare the PSB fraction in \firebox with results from recent observations \citep[][]{Wild2016,Rowlands2018a,Suess2022} at $z\simeq 0.5-1$. Note: the PSB fraction relative to the sample \citet{Suess2022} relies on non-uniformly selected objects from the Sloan Digital Sky Survey (SDSS) and here, it is thus not intended as a number density relative to that of either star-forming or all galaxies.}
    \label{fig:PSBs_fraction}
\end{figure}

\begin{figure*}
    \centering
    \includegraphics[width=\hsize]{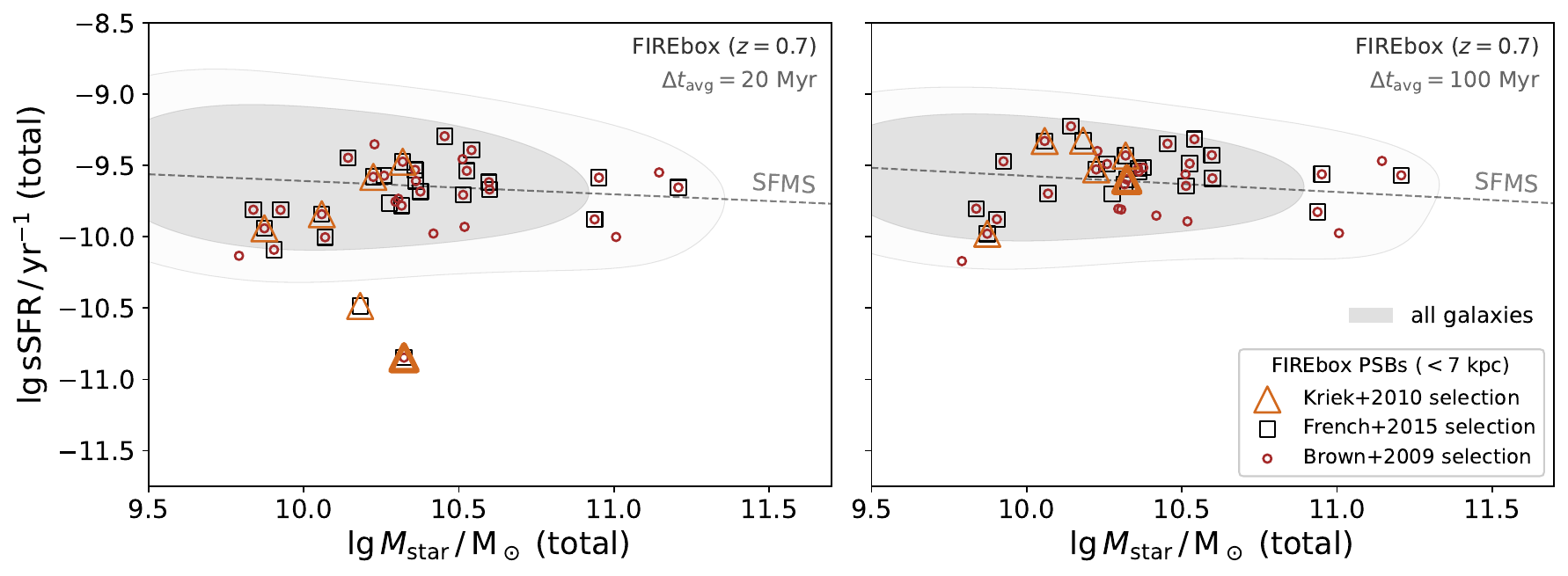}
    \includegraphics[width=\hsize]{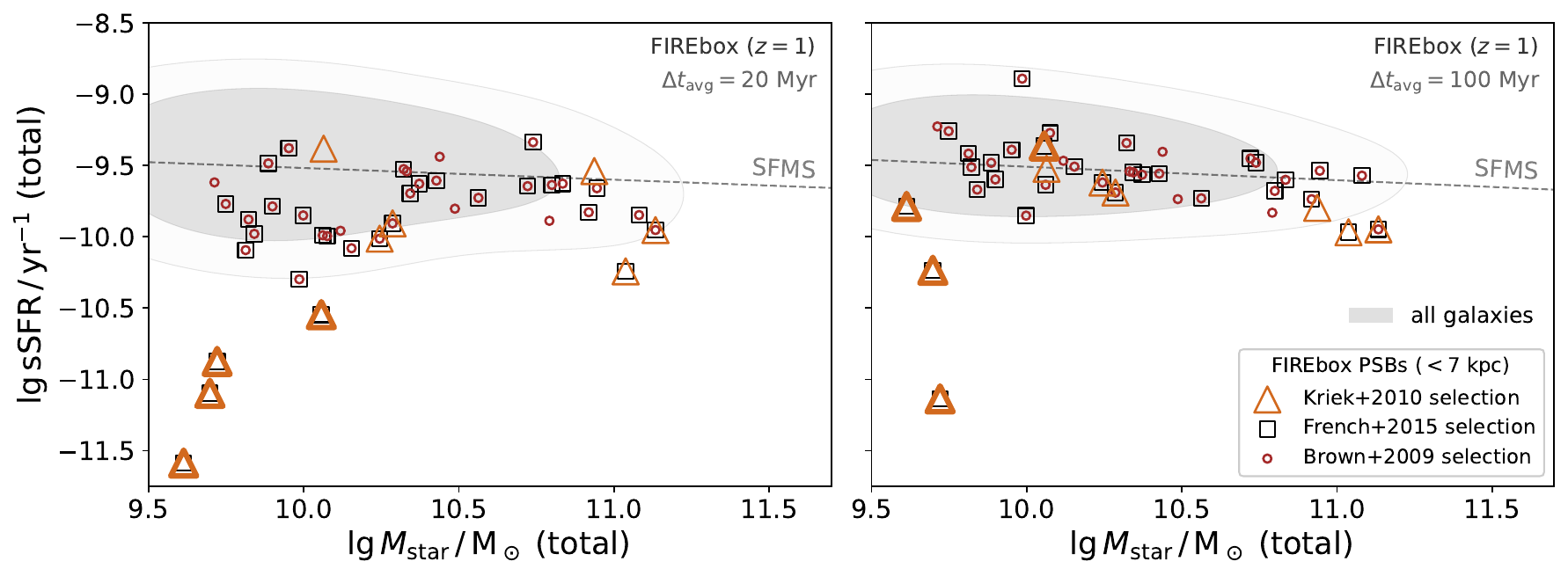}
    \caption{Total stellar mass ($M_{\rm star}$) and specific star-formation rate (sSFR) of $z=0.7$ (upper panels) and $z=1$ (bottom panels) \firebox galaxies with $M_{\rm star}\geq 3\times 10^{9}\Msun$. SFRs are computed using two different averaging times of 20 (left panels) and $100\Myr$ (right panels). We show PSBs selected with the fiducial criterion from \citet[][]{Kriek2010} (orange triangles), and measuring their photometry within an aperture of $7\kpc$. Spectroscopically-selected PSBs using selection criteria akin to those of \citet[][]{French2015} (black squares) and \citet[][]{Brown2009} (red circles) are also shown, again considering an aperture of $7\kpc$. The grey dashed lines represent the star forming main-sequence (SFMS) in \firebox. Grey, shaded contours show the 1-2$\sigma$ 2D distribution of all considered \firebox galaxies within this plane. The thick, orange triangles are photometry-selected (\citet{Kriek2010} criterion) PSBs with $\mathrm{sSFR}<3\times 10^{-11}~\mathrm{yr}^{-1}$ for $\tavg = 20\Myr$.}
    \label{fig:PSBs_SFMS}
\end{figure*}

\begin{figure*}
    \centering
    \includegraphics[width=\hsize]{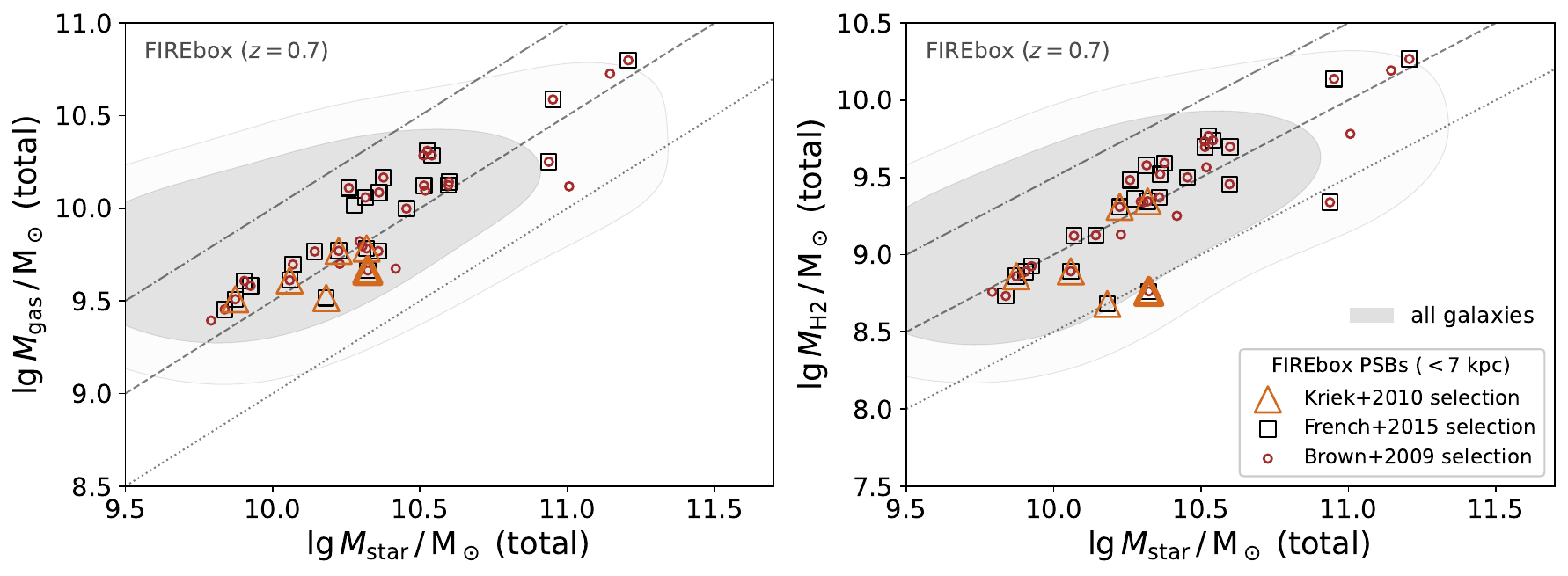}
    \includegraphics[width=\hsize]{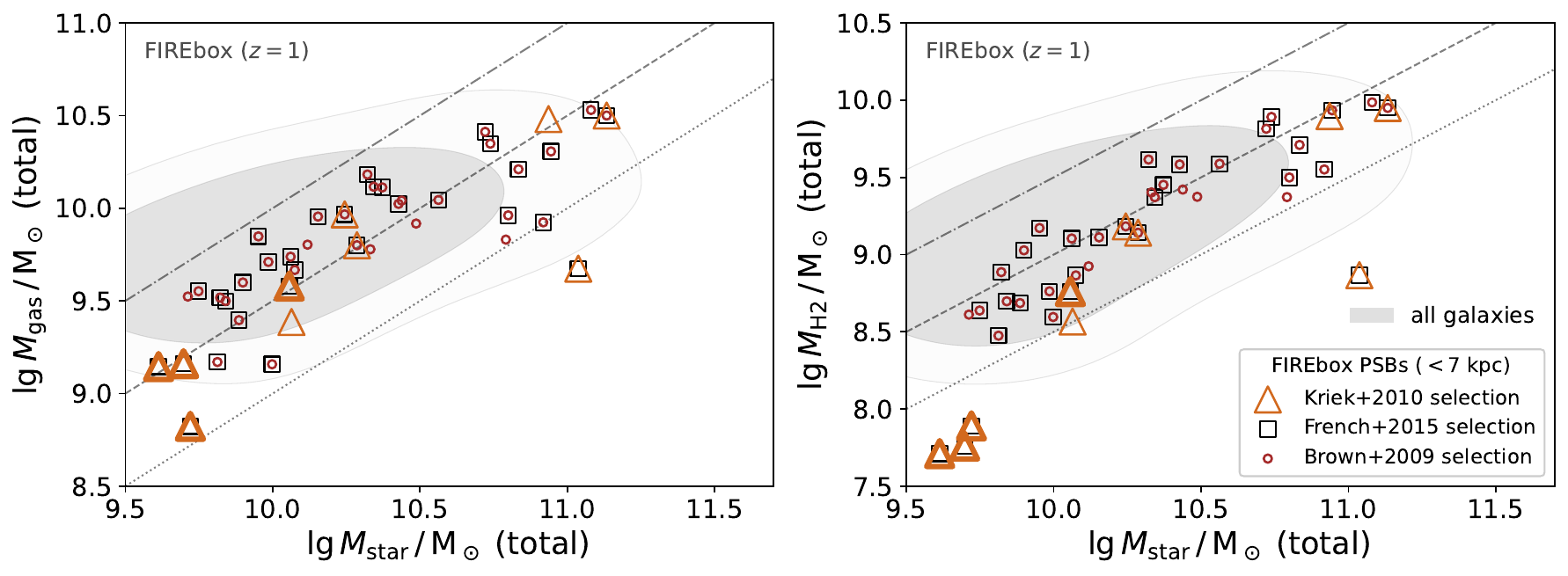}
    \caption{Total stellar mass ($M_{\rm star}$), total gas mass ($M_{\rm gas}$; left panel), and molecular gas mass ($M_{\rm H_2}$; right panel) of $z=0.7$ (upper panels) and $z=1$ (lower panels) \firebox galaxies with $M_{\rm star}\geq 3\times 10^{9}\Msun$. We show PSBs selected with the fiducial criterion from \citet[][]{Kriek2010} (orange triangles), and measuring their photometry within an aperture of $7\kpc$. Spectroscopically-selected PSBs using selection criteria akin to those of \citet[][]{French2015} (black squares) and \citet[][]{Brown2009} (red circles) are also shown, again considering an aperture of $7\kpc$. The grey dash-dotted, dashed, and dotted lines represent fixed values of the gas ratio $\mu_{\rm gas}\equiv M_{\rm gas}/M_{\rm star}=1,0.3,0.1$ (left panels) and molecular gas ratio $\mu_{\rm H_2}\equiv M_{\rm H_2}/M_{\rm star}=0.3,0.1,0.03$ (right panels), respectively. Grey, shaded contours show the 1-2$\sigma$ 2D distribution of all \firebox galaxies with $M_{\rm star}\geq 3\times 10^{9}\Msun$, within this plane, at the given redshift. The thick, orange triangles are photometry-selected (\citet{Kriek2010} criterion) with $\mathrm{sSFR}<3\times 10^{-11}~\mathrm{yr}^{-1}$ for $\tavg = 20\Myr$.}
    \label{fig:PSBs_H2}
\end{figure*}

\begin{figure}
    \centering
    \includegraphics[width=\hsize]{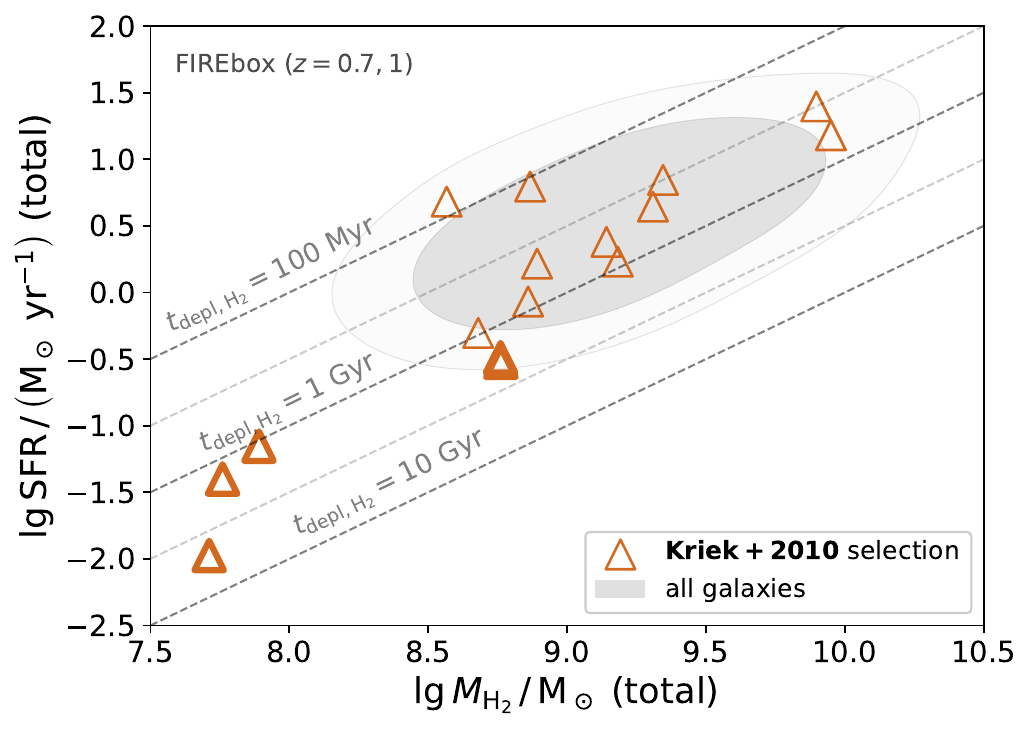}
    \caption{Total molecular gas mass ($M_{\rm H_2}$) and star-formation rate (SFR) of $z=0.7$ and $z=1$ galaxies in \firebox with $M_{\rm star}\geq 3\times 10^{9}\Msun$. SFRs are computed using an averaging time of $20\Myr$. For clarity, we only show PSBs selected with the fiducial criterion from \citet{Kriek2010}, measuring photometry within an aperture of $7\kpc$. The grey dashed lines represent fixed values of molecular-gas depletion times ($t_{\rm depl,\,H_2}\equiv M_{\rm H_2}/\mathrm{SFR}=0.1,0.3,1,3,10\Gyr$; between 0.1 and 1 Gyr, separated by 0.5 dex). Grey, shaded contours show the 1-2$\sigma$ 2D distribution of all considered \firebox galaxies within this plane. The thick, orange triangles are photometry-selected (\citet{Kriek2010} criterion) with $\mathrm{sSFR}<3\times 10^{-11}~\mathrm{yr}^{-1}$ for $\tavg = 20\Myr$. The majority of PSBs have an average molecular-gas depletion times of $t_{\rm depl,\,H_2}\simeq 400\Myr$, akin to that of star-forming galaxies in \firebox.}
    \label{fig:PSBs_tdepl}
\end{figure}

\begin{figure*}
    \centering
    \includegraphics[width=.329\hsize]{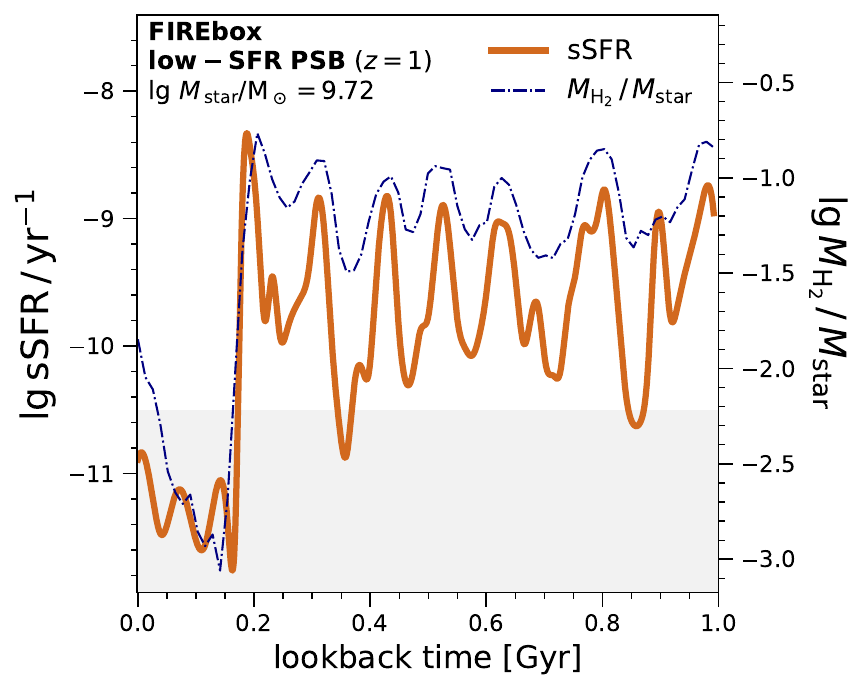}
    \includegraphics[width=.329\hsize]{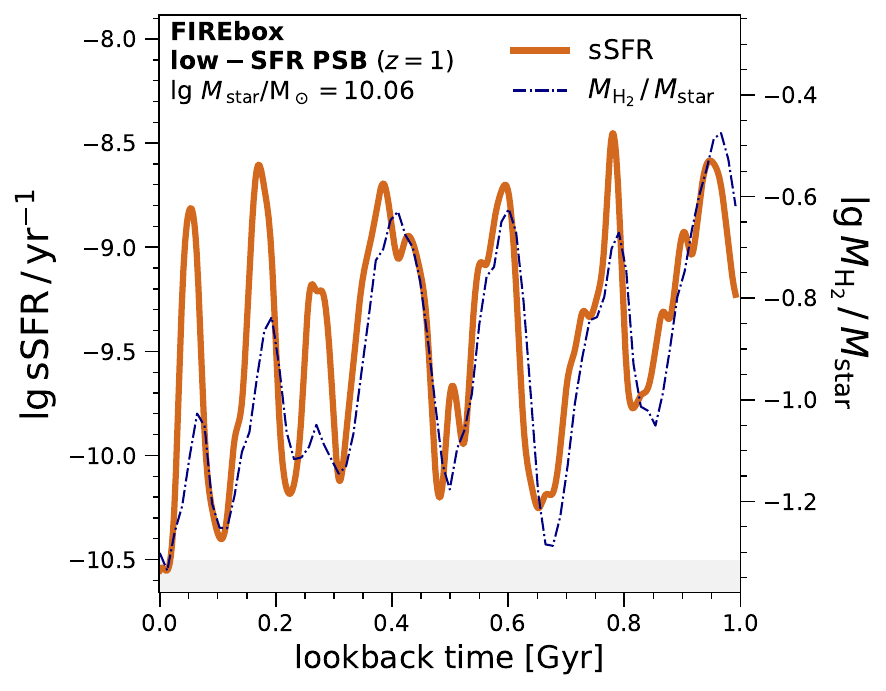}
    \includegraphics[width=.329\hsize]{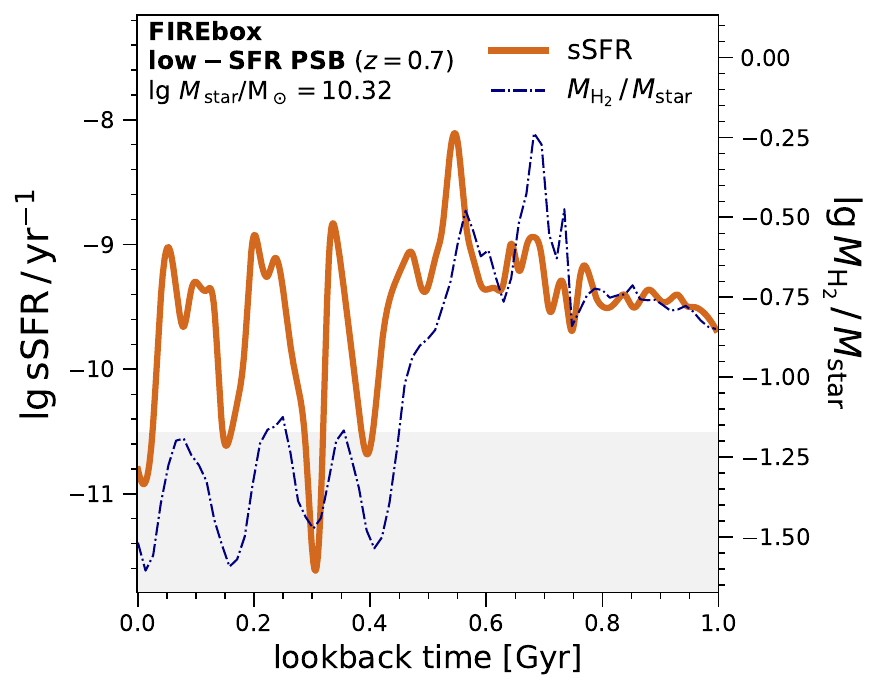}
    \includegraphics[width=.329\hsize]{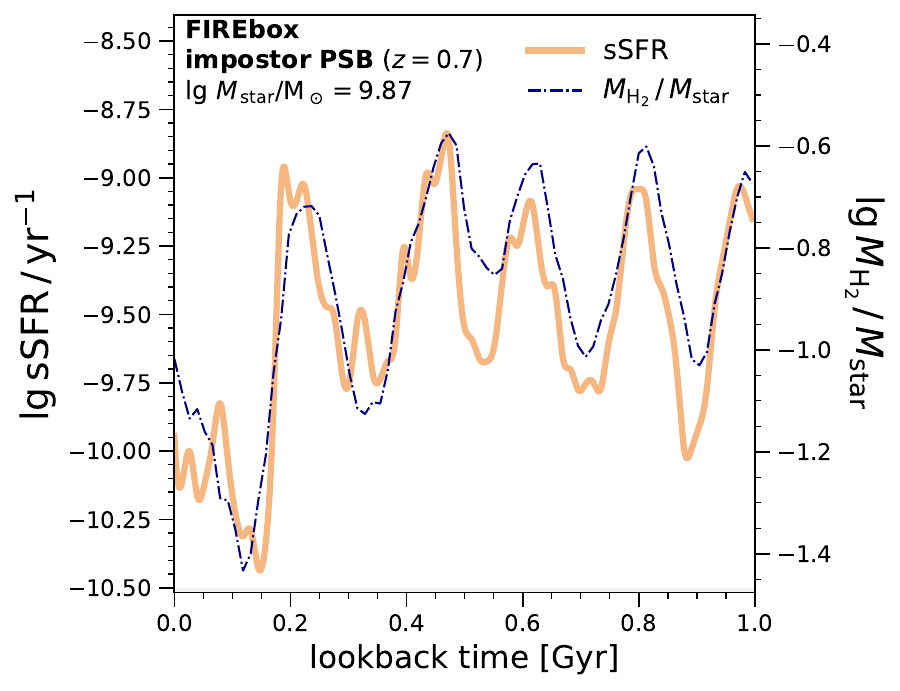}
    \includegraphics[width=.329\hsize]{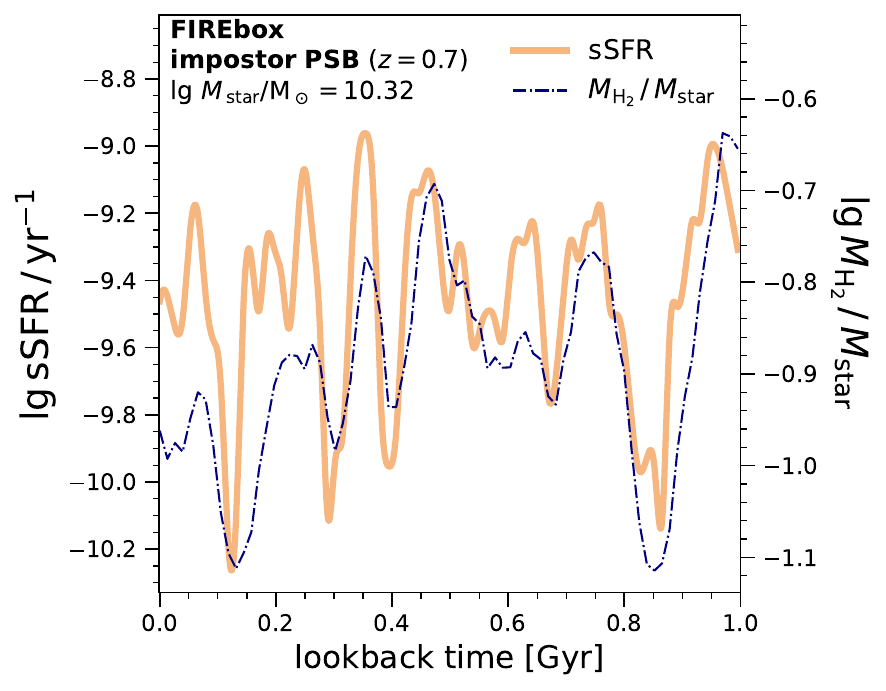}
    \includegraphics[width=.329\hsize]{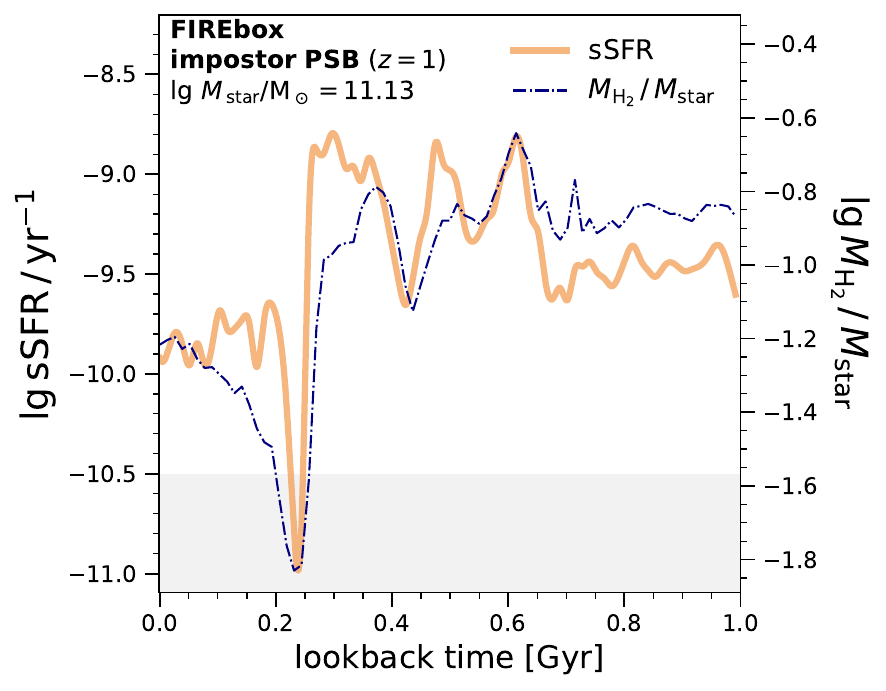}
    \includegraphics[width=.329\hsize]{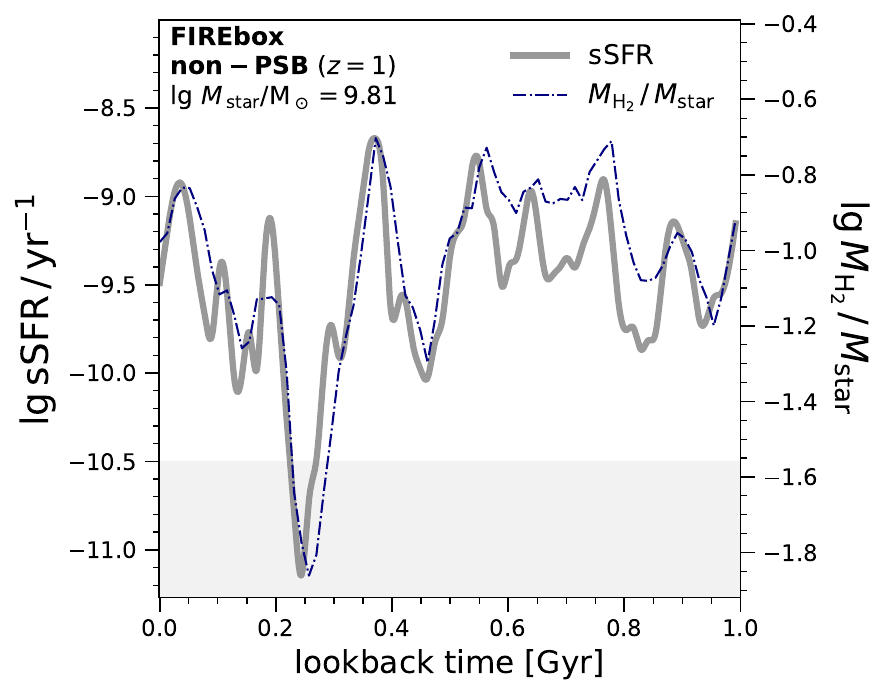}
    \includegraphics[width=.329\hsize]{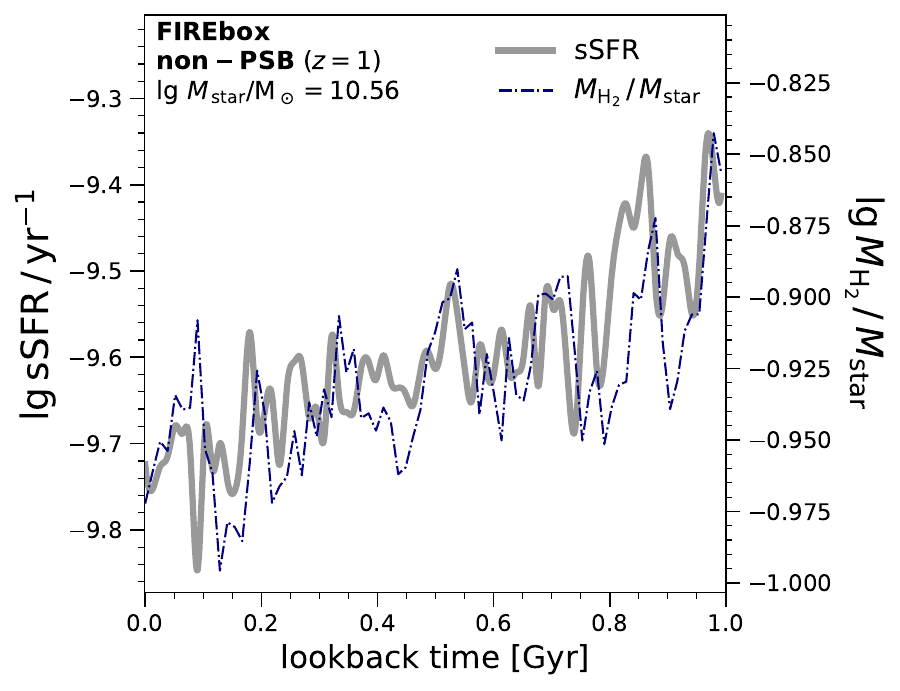}
    \includegraphics[width=.329\hsize]{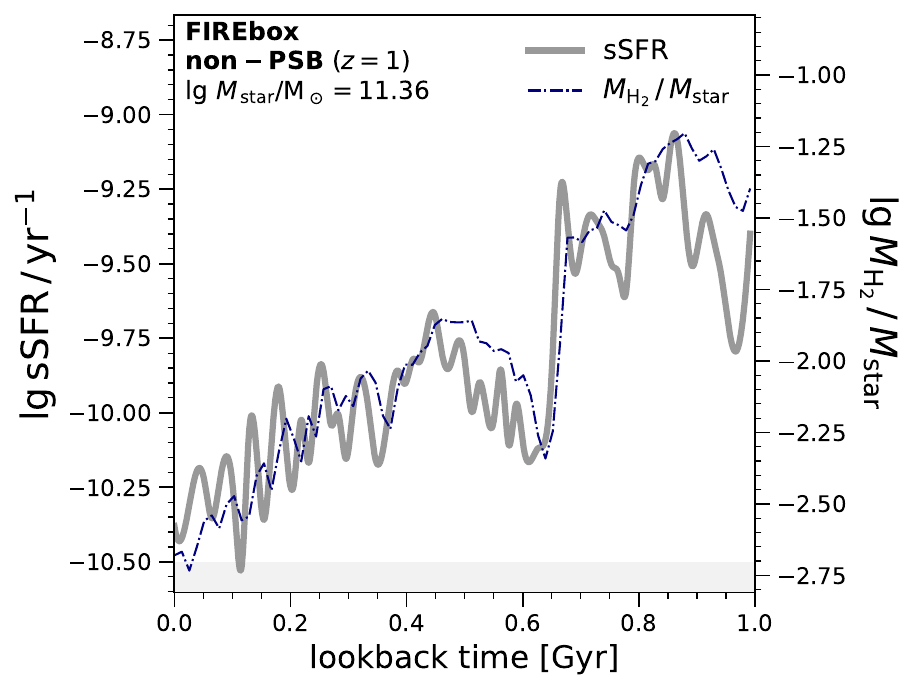}
    \caption{Examples of star-formation histories (SFH) of \firebox galaxies at $z=0.7$ and $z=1$ with $M_{\rm star}>3\times 10^9\Msun$. From left to right panels, galaxies have increasing stellar masses. We show the evolution of both specific SFR (sSFR; with $20\Myr$ averaging time; solid lines; left y-axes) and molecular hydrogen gas mass fraction ($M_{\rm H_2}/M_{\rm star}$; dash-dotted lines; right y-axes), over the past Gyr. We show the SFHs of PSBs that either have low SFRs (top row) or are currently star-forming (i.e. impostors; middle row) at the time when they are selected, as well as the SFHs of representative non-PSBs (bottom row). The shaded region marks values below the threshold $\mathrm{sSFR} = 3 \times 10^{-11}~\mathrm{yr}^{-1}$, which we use to identify low-SFR PSBs. Galaxies in \firebox have diverse SFHs, where the evolution of their SFR is tightly correlated with that of their molecular gas content. Most PSBs experienced a significant drop in their SFR and molecular gas fractions in the past $\lesssim 200\Myr$. Impostor PSBs in \firebox generally do show low-SFR periods but their SFR has since recovered to values that are typical of the galaxies close to the SFMS. Non-PSBs have a larger variety of SFH albeit they have not generally experienced a sudden drop in the past few hundred Myr.}
    \label{fig:PSBs_SFHs}
\end{figure*}

\begin{figure*}
    \centering
    \includegraphics[width=.9\hsize]{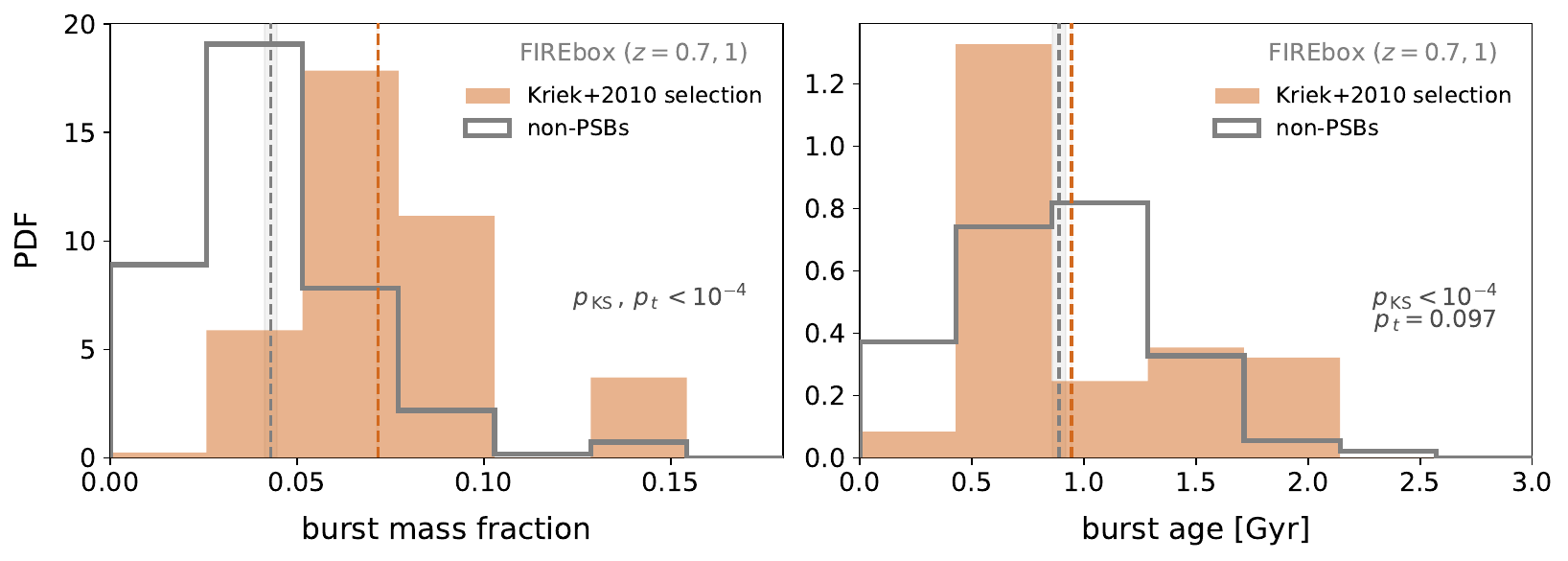}
    \caption{Estimated probability density function (PDF) for the burst mass fractions (left panel) and burst ages (right panel) of PSBs (\citet[][]{Kriek2010} selection; orange, filled histogram) and non-PSBs of similar stellar mass (grey line histogram) in \firebox ($z=0.7,1$; $M_{\rm star}>3\times 10^{9}\Msun$). Burst ages are defined as the time when these galaxies experienced their most recent major starburst event and burst mass fractions measure the fraction of their current stellar mass formed during the burst (see Section~\ref{sec:methods_burst} for details). Vertical, dashed lines show the estimated mean of the distributions. We report the p-values yielded by a Kolmogorov-Smirnov hypothesis test \citep[$p_{\rm KS}$;][]{Kolmogorov1933,Smirnov1939} and an independent (two-sample) t-test \citep[$p_{t}$;][]{Student1908}, to evaluate the differences between the distributions of PSBs and non-PSBs. On average, in \firebox, PSBs have larger burst mass fractions but similar burst ages than non-PSBs.}
    \label{fig:PSBs_burst_ages}
\end{figure*}

\begin{figure}
    \centering
    \includegraphics[width=\hsize]{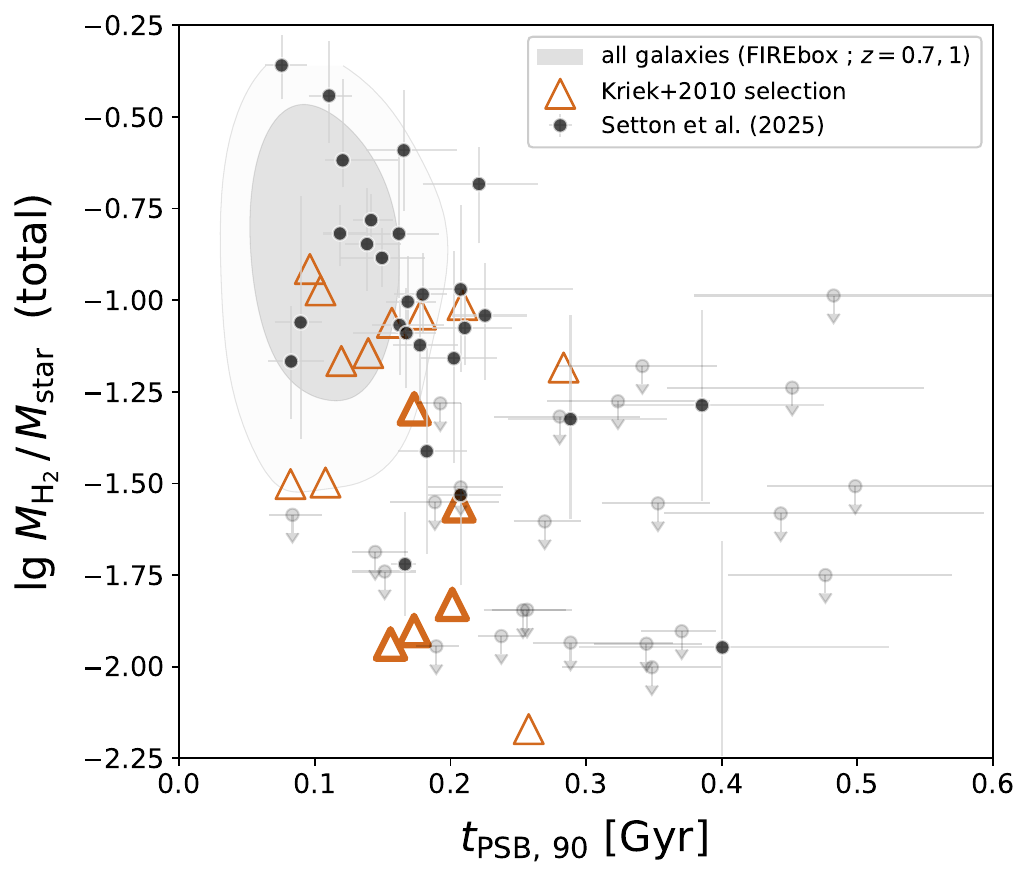}
    \caption{Molecular hydrogen gas fraction ($M_{\rm H2}/M_{\rm star}$) of \firebox galaxies in our sample as a function of the time since they assembled $90\percent$ of the stellar mass formed in the past Gyr ($t_{\rm PSB,90}$). Grey, shaded contours show the 1-2$\sigma$ 2D distribution of all \firebox galaxies at $z=0.7$ and $z=1$, with $M_{\rm star}\geq 3\times 10^{9}\Msun$. The orange triangles show \firebox PSBs selected following the fiducial criterion from \citet{Kriek2010}, based on their photometry within an aperture of $7\kpc$. The thick, orange triangles represent PSBs with $\mathrm{sSFR}<3\times 10^{-11}~\mathrm{yr}^{-1}$ for $\tavg = 20\Myr$. We also show recent estimates from ALMA observations of the \squiggle sample of PSB candidates at $z\simeq 0.7$ \citep[][]{Setton2025}. The error bars show the $1\sigma$ uncertainties on the \squiggle data points. Data points with increased transparency show upper limits on quantities on the y-axis. Galaxies in \firebox generally have molecular gas masses (and stellar masses) that are lower than the sample of observed galaxies in \squiggle, while the molecular gas fractions are in good agreement. PSBs in \firebox with $\mathrm{sSFR}<3\times 10^{-11}~\mathrm{yr}^{-1}$ have typical $\rm H2$ masses below the typical ALMA detection limits based on CO. Relatively molecular gas-rich PSBs in \firebox are almost exclusively impostors.}
    \label{fig:PSBs_H2_vs_ages}
\end{figure}

\begin{figure}
    \centering
    \includegraphics[width=\hsize]{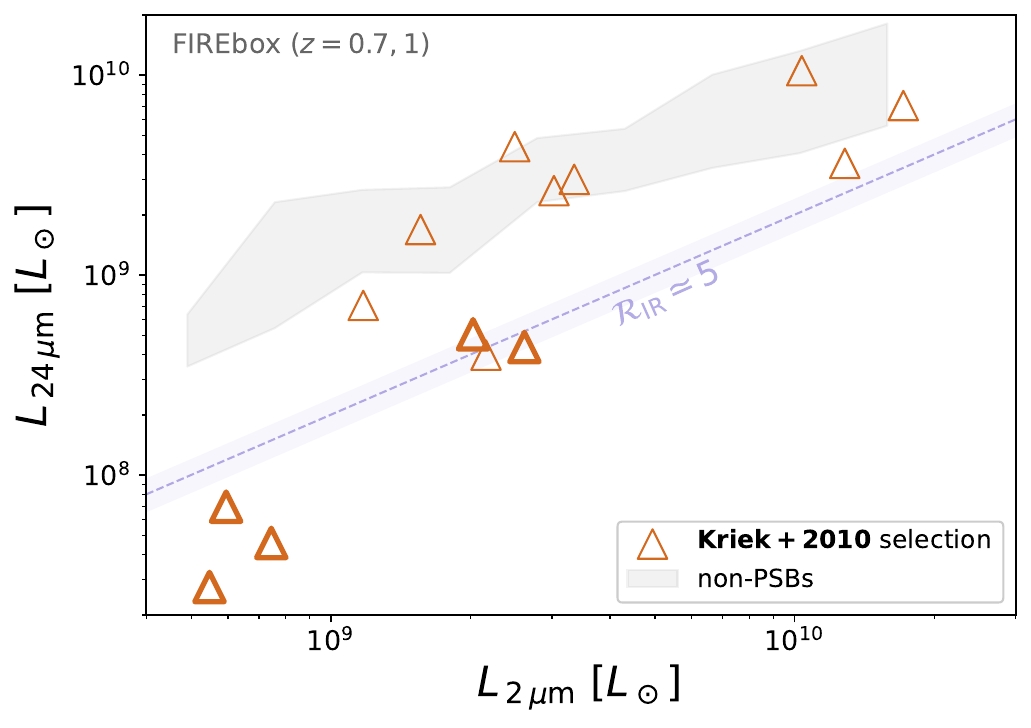}
    \caption{Mid-infrared (MIR) and near-infrared (NIR) luminosity ($\nu\,L_\nu$) of \firebox PSBs (orange triangles). The MIR and NIR luminosities are measured at the rest-frame wavelengths of $24$ $\mu$m and $2$ $\mu$m, respectively. The thicker markers are the low-SFR PSBs in \firebox (with $\mathrm{sSFR}<3\times 10^{-11}~\mathrm{yr}^{-1}$, for $\tavg = 20\Myr$), whereas thinner markers show star-forming PSBs (i.e. clear impostor PSBs). The dashed lines shows the optimal near-to-mid infrared ratio $\mathcal{R}_{\rm IR}=L_{2\mu\mathrm{m}}/L_{24\mu\mathrm{m}}\simeq 5$ we estimate (see main text for details) to separate star-forming PSBs from low-SFR PSBs. The latter, have lower MIR luminosities than other galaxies in \firebox with similar NIR luminosities. The grey shaded area shows the (16th to 84th percentiles) distribution of MIR luminosities at any given NIR luminosity, for a redshift- and mass-matched sample of non-PSBs in \firebox. On average, impostor PSBs have non-negligible infrared emission typical of star-forming galaxies and non-PSBs in \firebox, associated with on going star-formation.
    }
    \label{fig:PSBs_IR}
\end{figure}

\begin{figure*}
    \centering
    \includegraphics[width=.4975\hsize]{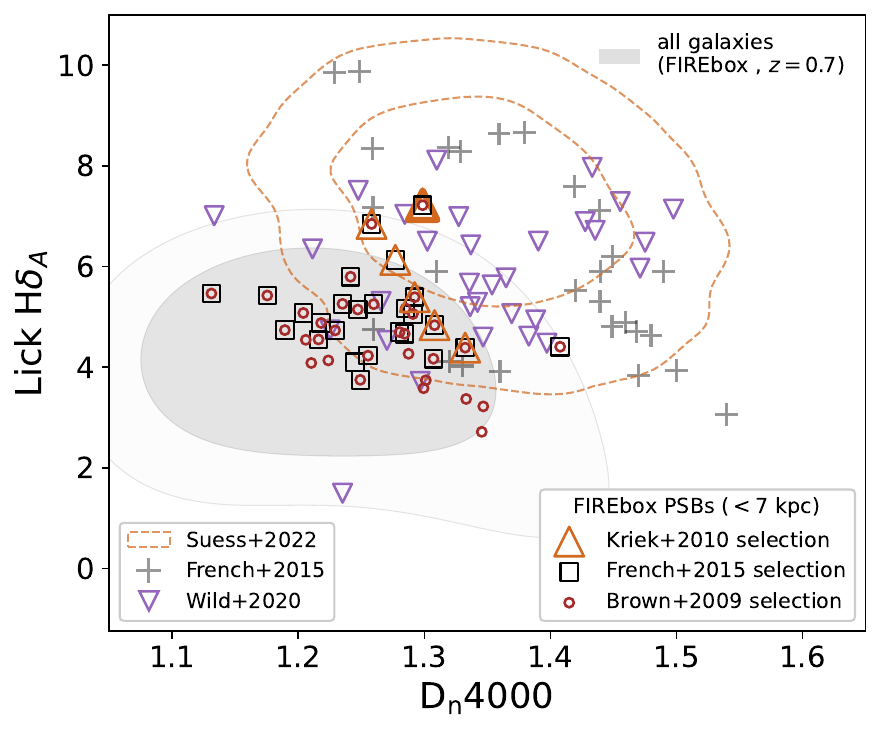}
    \includegraphics[width=.4975\hsize]{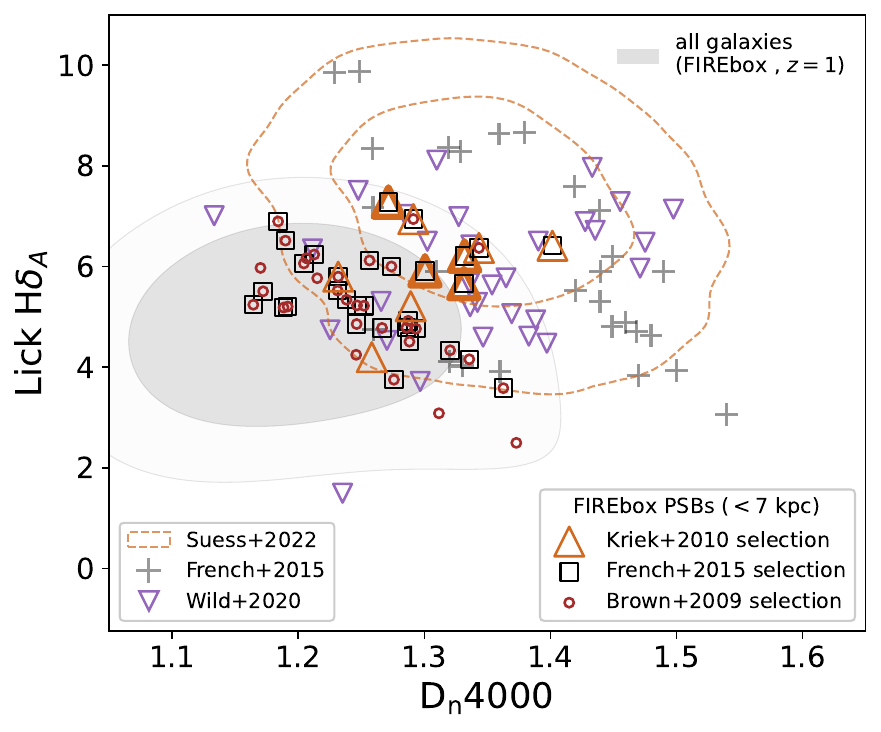}
    \caption{Median (over the considered projections) Lick H$\delta_{\rm A}$ and D$_{\rm n}4000$ for $z=0.7$ (left panel) and $z=1$ (right panel) \firebox galaxies with $M_{\rm star} > 3\times 10^9\Msun$ in comparison to observations of PSBs at $z\sim 0 - 1$. For \firebox, we show the distribution of all galaxies within this plane (black, dotted 1-2$\sigma$ contours) as well as the selected PSBs using different criteria: \citet[][]{Kriek2010} (orange triangles; fiducial criterion), \citet[][]{French2015} (black squares), and \citet[][]{Brown2009} (red circles). The thick, orange triangles represent \firebox PSBs selected with the \citet[][]{Kriek2010} criterion, with $\mathrm{sSFR}<3\times 10^{-11}~\mathrm{yr}^{-1}$ for $\tavg = 20\Myr$. All photometric and spectroscopic quantities for simulated galaxies are computed within an observational aperture of 7 kpc. Observational data include: \citet[][]{Suess2022} (orange, dashed 1-2$\sigma$ contours), \citet[][]{French2015} (black crosses), \citet[][]{Wild2020} (purple triangles). The Lick H$\delta_{\rm A}$ and D$_{\rm n}4000$ of \firebox PSBs are in reasonable agreement with observed values. In general, most \firebox galaxies have both lower Lick H$\delta_{\rm A}$ and D$_{\rm n}4000$ than \firebox PSBs and observed PSBs, possibly due to the younger stellar ages and lower stellar masses of the simulated sample.}
    \label{fig:PSBs_Hdelta}
\end{figure*}

\begin{figure}
    \centering
    \includegraphics[width=\hsize]{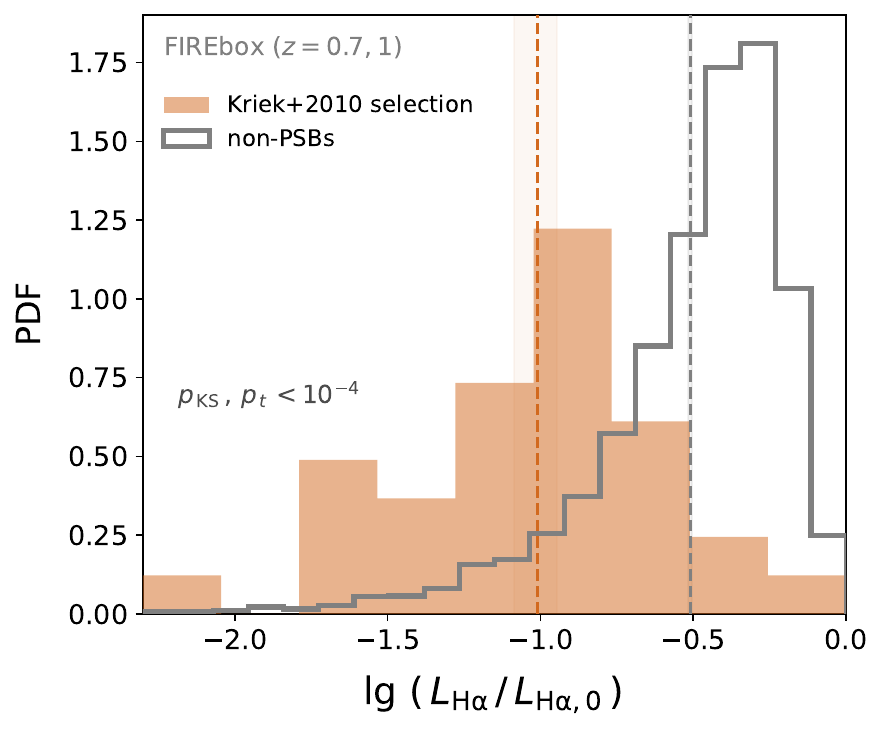}
    \caption{Estimated probability density function (PDF) for the ratio $L_{\rm H\alpha}/L_{\rm H\alpha,0}$ between the emergent and intrinsic $\rm H\alpha$ luminosities of PSBs (\citet[][]{Kriek2010} selection; orange, filled histogram) and a redshift- and mass-matched sample non-PSBs (grey line histogram) in \firebox ($z=0.7,1$; $M_{\rm star}>3\times 10^{9}\Msun$). Vertical, dashed lines show the estimated mean of the distributions. We report the p-values yielded by a Kolmogorov-Smirnov hypothesis test \citep[$p_{\rm KS}$;][]{Kolmogorov1933,Smirnov1939} and an independent (two-sample) t-test \citep[$p_{t}$;][]{Student1908}, to evaluate the differences between the distributions of PSBs and non-PSBs. On average, in \firebox, PSBs have lower $L_{\rm H\alpha}/L_{\rm H\alpha,0}$ than non-PSBs, indicating stronger effective attenuation of their $\rm H\alpha$ emission and suggesting that obscuration and projection effects may drive the emergence of impostor PSBs.}
    \label{fig:PSBs_Halpha}
\end{figure}

We identify PSBs in \firebox by following the selection criterion introduced in \citet{Kriek2010}, which relies on the flux in three synthetic medium-band optical filters \citep[$U_{\rm m}$, $B_{\rm m}$, and $V_{\rm m}$;][]{Kriek2010,Suess2022}. These filters are designed to isolate and quantify the strength of the $\sim 4000\AAA$ Balmer break and the slope of the SED at wavelengths longer than the break. Our initial sample, prior to PSB selection, consists of all galaxies in \firebox with stellar mass $M_{\rm star} \geq 3\times 10^{9}\Msun$ at the selected redshifts ($z=0.7$ and $z=1$), for a total of 310 galaxies (166 at $z=0.7$ and 144 at $z=1$). We only consider central galaxies and neglect satellite galaxies where star formation histories might be affected by environmental processes. We compute the $U_{\rm m}-B_{\rm m}$ and $B_{\rm m}-V_{\rm m}$ rest-frame colours (hereafter $\left(U-B\right)_{\rm rest}$ and $\left(B-V\right)_{\rm rest}$, respectively) within a number of fixed physical apertures of $1-10\kpc$ and within $0.1\,R_{\rm vir}$, for different viewing angles. PSBs have $\left(U-B\right)_{\rm rest} > 1$ and $\left(B-V\right)_{\rm rest} < 0.5$. Galaxies in our samples can be identified as a PSBs based on any of their projections and within any of the considered apertures. To compare results from different selection criteria based on spectroscopic properties, we also compute the Lick H$\delta_{\rm A}$ index \citep[][]{Worthey1994}{}{}, the D$_{n}4000$ index, measuring the strength of the Balmer break \citep[][]{Balogh1999}{}{}, and the equivalent widths (EW) for the Balmer-series spectral lines (e.g., H$\alpha\,6565$). Specifically, we select PSBs based on ($i$) the \citet{Brown2009} selection criterion, $\log\left[\mathrm{EW}\left(\mathrm{H}\alpha\right)\right] < 0.2\times \mathrm{Lick}~\mathrm{H}\delta_{\rm A}$ and $\mathrm{EW}\left(\mathrm{H}\delta_{\rm A}\right)>3$, and ($ii$) the \citet{French2015} selection criterion, $\mathrm{EW}\left(\mathrm{H}\alpha\right)<3$ (in emission) and $\mathrm{Lick}~\mathrm{H}\delta_{\rm A}>4$. We calculate the relevant spectroscopic quantities for every galaxy, seen under every projection, and within every considered aperture. To give robust estimates of the PSBs fraction and their average properties, we populate bootstrapped samples of $z=0.7$ and $z=1$ galaxies in \firebox with total stellar mass $M_{\rm star}\geq 3\times 10^{9}\Msun$, where each galaxy is viewed under a random projection among the 14 available.

In Figure~\ref{fig:PSBs_SED}, we show the average stacked spectral energy distribution (SED; rest-frame ultraviolet/optical) of the PSBs selected in \firebox in the fiducial sample, compared to those of the PSBs analysed in \citet{Suess2022} (\squiggle observational sample) and of a redshift- and mass-matched sample non-PSBs in \firebox. The SEDs shown in Figure~\ref{fig:PSBs_SED} were generated with \skirt using a wavelength grid with uniform linear spacing of $0.4~\AAA$ and launching $10^8$ virtual photon packets. However, to keep the fiducial \skirt simulations analysed in this work computationally feasible, we adopt a coarser wavelength grid and launch only $10^7$ photon packets (see Section~\ref{sec:methods_RT}). Overall, the shape of the spectrum of \firebox PSBs is in excellent agreement with that of observed PSBs in the \squiggle sample. The average SED of \firebox PSBs do not show any clear [OII]$\lambda 3726-\lambda 3729$ (doublet) emission compared to the \squiggle stacked data, likely due to a large AGN contamination in their \squiggle sample \citep[see][]{Greene2020,Suess2022}.

In summary, the fiducial PSB sample is populated by $z=0.7,1$ \firebox galaxies with $M_{\rm star}\geq 3\times 10^{9}\Msun$ that are selected as PSBs according to their $\left(U-B\right)_{\rm rest}$ and $\left(B-V\right)_{\rm rest}$ colours \citep[see][]{Kriek2010}, using a fixed fiducial aperture of physical radius of $7\kpc$, comparable to that of observational works \citep[e.g.,][]{Wild2016,Rowlands2018a}. Hereafter, we will alternatively refer to these PSBs selected in \firebox as ‘our' PSBs. Using apertures between $6-8\kpc$ to evaluate galaxy spectra and photometry does not change our main conclusions, however, using significantly smaller or larger apertures would definitely affect our general results. See, e.g., Figure~\ref{fig:PSBs_fraction} below for an example of the effect of changing apertures.

When computing the fractional contribution of PSBs among other galaxies (e.g., in Section~\ref{sec:discussion_impostor}), we consider all projections as independent and therefore every galaxy contributes to 14 entries in the final sample. To address average physical properties of PSBs (see, e.g., Figure~\ref{fig:PSBs_SFMS}) and the fraction of interacting systems (see, e.g., Figure~\ref{fig:PSBs_merger}), we will regard a galaxy as a PSB if it is selected as such based on its photometric properties in at least one of the 14 available projections. Changing the minimum number of projections (for instance, $3-5$) where a galaxy has to be selected as PSBs in order to contribute to their average properties only has a secondary effect on our conclusions.

\subsection{After-starburst galaxies and impostor PSBs}\label{sec:methods_ASBs}

We classify all \firebox galaxies based on their SFR history. Galaxies that experienced a significant (at least a factor of 4) drop in their current SFR since their most recent starburst event (defined in Section~\ref{sec:methods_burst}), are classified as \textit{after-starburst galaxies} (ASBs). We further define (temporarily) \textit{quenched after-starburst galaxies} (\text{q-ASBs}) as those ASBs that also have a present $\mathrm{sSFR}<3\times 10^{-11}~\mathrm{yr}^{-1}$, using a SFR averaging time of $\tavg=20\Myr$. These definitions are meant to mimic those employed by other numerical works aiming at selecting PSBs in cosmological volume simulations \citep[see, e.g.,][]{Davis2019}. In contrast to these studies, our approach is to select PSBs based on their observational properties \citep[as in, e.g.,][]{Kriek2010,Suess2022} and evaluate whether or not the galaxies in our PSB samples (see Section~\ref{sec:methods_samples} above) have properties akin to those expected for real PSBs in the Universe.

Some galaxies that are selected as PSBs either did not experience a recent, significant reduction in their SFR, or have current SFRs that are not low enough to be considered quiescent ($\mathrm{sSFR}<3\times 10^{-11}~\mathrm{yr}^{-1}$). In other words, PSBs that are not classified as \text{q-ASBs} are clearly \textit{impostors}. Therefore, the fraction $\Fim$ of clear PSB impostors among the PSB population in \firebox is given by:
\begin{equation*}
    \mathcal{P}\left(\text{Im}\lvert\text{PSB}\right) \,=\, 1-\mathcal{P}\left(\text{q-ASB}\lvert\text{PSB}\right) \,\equiv\, \Fim
    ~.
\end{equation*}

In Table~\ref{tab:defs}, we summarise the relevant classes of galaxies used in this work, and report their corresponding fractional contributions to the entire galaxy population in \firebox and to the population of PSBs selected in \firebox.

\section{Results}\label{sec:results}

\subsection{Post-starburst fraction}\label{sec:results_sample}

Figure~\ref{fig:PSBs_impostor_frac} illustrates the resulting fraction of ASBs and \text{q-ASBs} (see Section~\ref{sec:methods_ASBs} for the corresponding definitions) among the PSBs in \firebox (left panel) and the fraction of PSBs and \text{q-ASBs} among ASBs (right panel). We refer the reader to Table~\ref{tab:defs} for a summary of the relevant classes of galaxies used in this work (such as PSBs, ASBs, \text{q-ASBs}) and corresponding fractional contributions to the entire galaxy population in \firebox. In the fiducial sample (\citet{Kriek2010} selection criterion, aperture f $7\kpc$), PSBs represent $3.1\percent$ of all \firebox galaxies at $z=0.7$ and $z=1$, with stellar masses $M_{\rm star}>3\times 10^{9}\Msun$. Restricting the sample to star-forming galaxies (i.e. with $\mathrm{sSFR}>3\times 10^{-11}~\mathrm{yr}^{-1}$), this fraction reduces to $2.2\percent$. About $77.6\percent$ of PSBs are ASBs and only $31.8\percent$ of PSBs are also \text{q-ASBs} (and thus defined as \true PSBs in \firebox). Among all galaxies, about $37.4\percent$ are ASBs, while only $2.6\percent$ are \text{q-ASBs}. Note that, by construction, all \text{q-ASBs} are also ASBs (i.e. $\mathcal{P}\left(\text{ASB}\lvert\text{q-ASB}\right) = 1$). Non-PSBs (i.e. galaxies that are not selected as PSBs in a given projection) are $96.9\percent$ of all \firebox galaxies in our sample, increasing to a contribution of about $98\percent$ among star-forming galaxies. About $36.1\percent$ of non-PSBs are ASBs and only $1.6\percent$ of them are also \text{q-ASBs}. 

Figure~\ref{fig:PSBs_fraction} shows the fraction of PSBs among all \firebox galaxies at $z=0.7$ and $z=1$, with stellar masses $M_{\rm star}>3\times 10^{9}\Msun$, as a function of physical observational aperture from the galaxy centre. Using the \citet{Kriek2010} selection criterion, based on rest-frame optical colours, the PSB fraction in \firebox varies from $2$ to about $4\percent$, with shrinking aperture radius from $12$ to $1\kpc$. We also show the PSB fractions relative to the two selection criteria based on spectroscopic properties \citet[][]{Brown2009,French2015,French2018} (see Section~\ref{sec:methods_samples} for further details on sample selection). For apertures larger than about $6\kpc$, the different criteria give PSB fractions that span from $\lesssim 7.5\percent$ for the \citet{French2015} selection criterion to about $2\percent$ for the \citet{Kriek2010} one (fiducial sample), independent on the chosen aperture. We report uncertainties on the median PSB fractions in \firebox, arising from both the relatively small size of the sample and the differences in the colours/spectra between different projections of the same galaxy. \textit{Overall, PSB fractions in \firebox are in good agreement with observational estimates} \citep[e.g.,][]{Wild2016,Rowlands2018a,Suess2022}. The dependence on projection and aperture size suggest that the role of dust content and geometry may significantly affect the observational properties of PSBs in \firebox.

In the present work, PSBs \firebox are classified as impostors if they are either actively star-forming (with $\mathrm{sSFR}>3\times 10^{-11}~\mathrm{yr}^{-1}$) or did not experience a recent significant reduction in their star formation (see Section~\ref{sec:methods_ASBs}). We quantify their fraction as that of PSBs that are not \text{q-ASBs}, that is about $68.2\percent$ (${P}\left(\text{Im}\lvert\text{PSB}\right) = {P}\left(\text{q-ASB}\lvert\text{PSB}\right)\equiv \FimFB= 0.682$). The emergence of impostor PSBs in our sample may result from non-standard dust attenuation laws and obscuration of star-forming regions \citep[see, e.g.,][]{Baron2022,Baron2023}, as well as aperture and projection effects. However, dust obscuration is unlikely the sole contributor towards explaining the difference between impostors and true PSBs \citep[see, e.g.,][]{Zhu2025}. Furthermore, in \firebox, PSBs could over-represent a population of low-mass galaxies with bursty and periodic star-formation history akin to that of breathing galaxies presented in \citet{Cenci2024a}.

Using Bayes' theorem, we find that the fraction of \text{q-ASBs} among \firebox PSBs is about $38.1\percent$ (i.e. ${P}\left(\text{PSB}\lvert\text{q-ASB}\right)=0.381$), whereas the probability ${P}\left(\text{q-ASB}\lvert\text{PSB}\right)$ of a galaxy being a \text{q-ASB}, given that it is selected as PSB in \firebox is lower, about $31.8\percent$ (see above). In other words, \firebox galaxies that experienced a significant drop in their SFR following a recent starburst and now have a low SFR are often displaying properties similar to those of PSBs selected in \firebox \citep[cf.,][]{Davis2019}. Conversely, \firebox PSBs poorly represent the population of (recently, and often temporarily) quenched after-starburst galaxies.

\subsection{Impostor PSBs on the star-forming main sequence}\label{sec:results_properties_SFR}

Figure~\ref{fig:PSBs_SFMS} shows the distribution of specific SFRs of \firebox galaxies at $z=0.7$ and $z=1$, as a function of their stellar mass, for two different averaging times for star formation ($\tavg=20,100\Myr$). \textit{Many of the PSBs in \firebox are clear impostors, lying within the characteristic scatter of the star-forming main sequence} (SFMS; calculated in \firebox), for both $\tavg=20,100\Myr$. However, a number of PSBs do display lower levels of star formation given their stellar mass and redshift, well below the SFMS. The fraction of low-SFR PSBs in \firebox is larger at higher redshifts and lower stellar masses. Furthermore, this fraction increases when considering $\tavg = 20\Myr$, compared to $\tavg = 100\Myr$, indicating that these galaxies have experienced a significant drop (about 1-1.5 dex) in their SFR in their recent past ($\lesssim 100\Myr$). In fact, PSBs with particularly low SFRs, below $\mathrm{sSFR}<3\times 10^{-11}~\mathrm{yr}^{-1}$, are all selected as \text{q-ASBs} (see Table~\ref{tab:defs} for details on definitions). The different selection criteria used in this work yield different fractions of PSBs, with spectroscopic criteria preferentially selecting star-forming galaxies. With the \citet{French2015} criterion does select all of the low-SFR PSBs identified with the \citet{Kriek2010}, yielding an overall larger fraction of selected PSBs but a similar fraction of low-SFR PSBs, and indicating the photometry-PSBs selected in \firebox do exhibit faint $\rm H\alpha$ emission. On the other hand, the \citet{Brown2009} almost exclusively selects impostors. In this work, we primarily discuss the results of our analysis on the fiducial PSB sample, populated using the \citet{Kriek2010} selection criterion. We defer the study of the effect of employing  different PSB selection criteria to a future work, with a larger sample of simulated galaxies.

\subsection{Molecular gas-rich impostor PSBs}\label{sec:results_properties_gas}
Figure~\ref{fig:PSBs_H2} shows the total gas and molecular gas content of \firebox galaxies at $z=0.7$ and $z=1$, as a function of their stellar mass. The majority PSBs have both total gas and molecular gas masses that are consistent (within 2$\sigma$) with those of star-forming galaxies in \firebox, with a similar stellar mass. This further support the interpretation of these galaxies as impostor PSBs in the simulation. A few PSBs with the lowest sSFR (for $\tavg = 20\Myr$; see Figure~\ref{fig:PSBs_SFMS}) do host a non-negligible amount of both total and molecular gas, indicating that, although their star formation activity may have dropped in the last $<100\Myr$, they still host relatively low-density \citep[i.e. not eligible for star-formation in our model; see also][]{Cenci2024a} molecular gas. However, those PSBs with a consistently low average sSFR across the past $100\Myr$ also have relatively low molecular gas mass fractions compared to the bulk of all \firebox galaxies ($\mu_{\rm H_2}\equiv M_{\rm H_2}/M_{\rm star}\lesssim 0.01$). These objects represent the most dramatic cases of temporarily quenched PSBs in our sample. 

Figure~\ref{fig:PSBs_tdepl} shows the molecular gas mass and SFR (for $\tavg = 20\Myr$) of the \firebox galaxies at $z=0.7$ and $z=1$ considered in this work. PSBs in \firebox have average molecular-gas depletion times (i.e. the ratio $t_{\rm depl,\,H_2}\equiv M_{\rm H_2}/\mathrm{SFR}$) of about $700\Myr$. PSBs with the lowest SFRs (specifically, with $\mathrm{sSFR}>3\times 10^{-11}~\mathrm{yr}^{-1}$) and molecular gas fractions ($\mu_{\rm H_2}\lesssim 0.01$) result in longer $t_{\rm depl,\,H_2}\simeq 2\Gyr$, on average. Star-forming PSBs, with SFRs comparable to those of non-PSBs with similar stellar mass, also have similar molecular-gas depletion times of about $400\Myr$.

To summarise, \textit{many of the PSBs selected in \firebox, based on their photometry and spectroscopic properties, are actively star-forming host abundant total and molecular gas reservoirs, and have molecular depletion times} (or, equivalently, star-forming efficiencies) \textit{similar to those of star-forming non-PSBs}.

\subsection{Star-formation histories}\label{sec:results_properties_SFR}

Figure~\ref{fig:PSBs_SFHs} shows examples of star-formation histories (SFH) of PSBs and non-PSBs in \firebox. Specifically, we show the evolution of both specific SFR and molecular hydrogen gas mass fraction, over the past Gyr. In general, galaxies in \firebox have diverse SFHs. The evolution of the SFR of \firebox galaxies is tightly correlated with that of their molecular gas content, where increasing $M_{\rm H_2}$ are typically associated with proportionally increasing SFRs and significantly reduced $M_{\rm H_2}$ correspond to the drop in SFR due to gas consumption and stellar feedback following starburst events. Consequently, after major starbursts, both the SFR and $M_{\rm H_2}$ of galaxies are drastically reduced.

The PSBs with low star-formation activity at the time when they are selected (top panels in Figure~\ref{fig:PSBs_SFHs}) are characterized by a recent, intense starburst event in the past $\lesssim 200\Myr$, resulting in a significant drop in their sSFR. During this starburst event, a large fraction of the available gas reservoir in these galaxies is consumed, ejected, and diffused, temporarily halting star formation. This behaviour is akin to that of ‘breathing' galaxies, where stellar feedback periodically displaces large amounts of gas, especially in low-mass galaxies \citep[e.g.,][]{Stinson2007,Muratov2015,Christensen2016,El-Badry2016,Angles-Alcazar2017,Pandya2021,Cenci2024a}. This starburst-induced quenching episode is only temporary, and the galaxies will eventually re-establish a star-formation activity close to that of galaxies on the star-forming main-sequence (SFMS). The recent evolution of low-SFR PSBs is thus consistent with that of low-mass, non-interacting starburst galaxies studied in \citet{Cenci2024a}. The SFH and properties of temporarily quenched galaxies in our sample that exhibit PSB-like photometric features are also similar to rejuvenating galaxies observed at high-redshift \citep[$z>5$; see, e.g.,][]{Witten2024}. We posit that our simulated sample of galaxies clearly includes a subset of galaxies cycling between bursty and quiescent phases, which may be over-represented in \firebox relative to the real Universe. Their rapid variability can exaggerate discrepancies between instantaneous and time-averaged SFR indicators, complicating the direct comparisons with observational samples and reproducing features typical of real PSBs.

Over the past Gyr, the overall SFHs of star-forming PSBs (middle panels in Figure~\ref{fig:PSBs_SFHs}) are akin to those of low-SFR PSBs, with periodic bursts of star formation followed by periods with lower SFR and reduced molecular gas fractions. In some cases, these impostor PSBs in the simulation are selected soon ($\lesssim 100\Myr$) after the time when their sSFR reached sufficiently low values to be classified as \true PSBs in \firebox. Alternatively, some impostor PSBs are instead observed a few hundreds Myr after the most recent major drop in SFR but their SFR could not increase to values similar to their average SFR over the past Gyr. In general, mild and recent rejuvenation could allow for galaxies to still be selected as PSBs when only examining their optical photometric properties. Massive galaxies (see right panels) exhibit a bursty, oscillating SFH only in the recent past $\lesssim 500\Myr$, triggered by a major galaxy interaction further in the past (see Section~\ref{sec:results_int} for results on the role of interactions in PSBs).

Non-PSBs in \firebox (bottom panels in Figure~\ref{fig:PSBs_SFHs}) have SFHs that are generally less bursty in the last Gyr, compared to PSBs. However, some non-PSBs do have SFHs that resemble those of impostor PSBs, with a major (albeit temporary) drop in SFR and molecular gas fractions happening in the past few hundreds Myr and prior to (recent) rejuvenation bringing them back to star formation levels that are typical of normal star-forming galaxies.

\subsection{Burst properties}\label{sec:results_burst}
In Figure~\ref{fig:PSBs_burst_ages}, we show the estimated probability density function (PDF) for the burst ages and burst mass fractions (see definitions in Section~\ref{sec:methods_burst}) of PSBs and non-PSBs of similar mass in \firebox \citep[at $z=0.7$ and $z=1$; fiducial PSB selection criterion from][]{Kriek2010}. We populate a redshift- and mass-matched ($\Delta \lg\,M_{\rm star}<0.1~\mathrm{dex}$) random control sample of unique non-PSBs for the sample shown in Figure~\ref{fig:PSBs_burst_ages}. PSBs contribute to the distribution with a weight that is given by the fraction of their 14 available projections that satisfy the fiducial PSB selection criterion. Projections that are not selected as PSBs contribute to the distribution of non-PSBs instead. On average, PSBs in \firebox have larger burst mass fractions but similar burst ages than non-PSBs with similar stellar masses. In \firebox, both classes of galaxies have burst mass fractions ranging from $<1\percent$ to about $15\percent$ and burst ages up to about $2.5\Gyr$. On average, PSBs (non-PSBs) in \firebox form about $6.6\percent$ ($4.3\percent$) of their stellar mass during their most recent starburst event, that took place, about $\lesssim 1\Gyr$ in their past. Burst ages of \firebox PSBs are generally much larger than the time since quenching reported for PSBs in the \squiggle sample \citep[$\lesssim 0.4\Gyr$;][]{Suess2022}. At the same time, the burst mass fractions of \firebox PSBs are much smaller than those estimated by \citet{Suess2022}. This discrepancy can originate from differences in the methods, especially in the time window during which we consider stars to have formed in the same starburst event (in this work, this is about $150\Myr$ around the peak of star formation; see Section~\ref{sec:methods_burst}). Furthermore, \firebox PSBs are generally less massive than \squiggle PSBs, and might therefore have experienced milder starbursts. However, in \firebox, we do not see any significant trend of either burst mass fraction or burst age with stellar mass, in the mass and redshift range explored in this work.

In Figure~\ref{fig:PSBs_H2_vs_ages}, we show the molecular hydrogen gas fractions ($M_{\rm H_2}/M_{\rm star}$) of the \firebox galaxies in our sample as a function of the time when they assembled $90\percent$ of the stellar mass that formed in the past Gyr ($t_{\rm PSB,90}$; see Section~\ref{sec:methods_burst}). We compare our results to observational data, using estimated molecular gas contents from ALMA observations of galaxies in the \squiggle sample of PSB candidates at $z\simeq 0.7$ \citep[][]{Setton2025}. PSBs in \firebox have molecular gas masses that are generally smaller that those estimated for \squiggle PSBs \citep[][; see also Figure~\ref{fig:PSBs_H2}]{Setton2025}, owing to the latter targeting, on average, more massive galaxies. The average $\mathrm{H_2}$ masses among the PSBs selected in \firebox (about $5\times 10^8\Msun$) is likely below current detection limits at $z\gtrsim 0.7$. On average, the $\rm{H}_2$ fractions and $t_{\rm PSB,90}$ of PSBs in \firebox are in good agreement with observations with confirmed $\rm{H}_2$ detections. The non-detection of observed PSBs with a relatively long $t_{\rm PSB,90}\gtrsim 200\Myr$ suggests that they have very small $\rm{H}_2$ contents, with a potential correlation between the molecular gas mass and $t_{\rm PSB,90}$ of PSBs. These observations may reveal an evolutionary sequence as galaxies undergo a PSB phase, where their molecular gas mass is either depleted or ejected as they age \citep[e.g.,][]{Bezanson2022}. However, it is also possible that data are sampled from two distinct populations with short and long estimated $t_{\rm PSB,90}$. In general, galaxies with $t_{\rm PSB,90}\lesssim 100\Myr$ are consistent with being actively star-forming and had a constant average star-formation history over their past Gyr. PSBs in \firebox have estimated $t_{\rm PSB,90}\sim 100-200\Myr$, possibly indicating a decline in their recent ($<1\Gyr$) star formation (as in the case of ASBs, by definition). We suggest that, \textit{many of the observed PSBs with significant CO detections are impostors, with significant on-going star formation}. On the other hand, PSBs in \firebox with low SFRs have average molecular gas fractions consistent with those of observed PSBs with too small $\rm{H}_2$ masses to be detected in CO with ALMA. Therefore, in our picture, objects with low molecular gas fractions are more likely to be true, rapidly quenched PSBs, that could rejuvenate in their near future. Galaxies with a long $t_{\rm PSB,90}$ have formed a negligible fraction of their mass in recent times compared to their activity in the past Gyr, consistently with the theoretical expectation for the nature of PSBs. However, \firebox galaxies generally yield $t_{\rm PSB,90}\lesssim 300\Myr$, possibly due to the under-predicted populations of long-term quenched galaxies, likely related to the absence of black hole physics in the simulation. See Section~\ref{sec:discussion_agn} for a discussion of the possible role of black hole feedback in driving PSBs.

\subsection{Observational properties of PSBs}\label{sec:results_properties_obs}
In Figure~\ref{fig:PSBs_IR}, we show the mid-infrared (MIR) and near-infrared (NIR) luminosities for the PSBs selected in \firebox. Specifically, luminosities are computed as the total rest-frame emission ($\nu\,L_\nu$) of the target galaxies at a given wavelength. The MIR luminosities are computed at the rest-frame wavelength of $24$ $\mu$m ($L_{24\mu\mathrm{m}}$) and NIR luminosities at the rest-frame wavelength of $2$ $\mu$m ($L_{2\mu\mathrm{m}}$). The MIR emission at rest-frame $24$ $\mu$m is generally a proxy for SFR \citep[see, e.g.,][]{Calzetti2013}, while the NIR $2$ $\mu$m emission gives an approximated estimate for stellar mass \citep[e.g.,][]{Conroy2013}. In general, PSBs with low SFRs (specifically, with $\mathrm{sSFR}<3\times 10^{-11}~\mathrm{yr}^{-1}$ for $\tavg = 20\Myr$) have lower $24$ $\mu$m luminosities than star-forming, impostor PSBs in \firebox with similar NIR luminosities, due to their lower intrinsic SFR at fixed stellar mass. By contrast, impostor PSBs and non-PSB star-forming galaxies have similar MIR luminosities at a given NIR luminosity. The near-to-mid infrared ratio may thus be used to distinguish observationally between truly low-SFR PSBs and impostor PSB. Using logistic regression on our fiducial PSB sample in \firebox, we determine that the optimal near-to-mid infrared ratio $\mathcal{R}_{\rm IR}$ to separate low-SFR PSBs and star-forming PSBs. Specifically, by weighting each galaxy by the number of projections in which it is selected as a PSB, logistic regression yields the best linear separation for:
\begin{equation}
    \mathcal{R}_{\rm IR} \equiv \frac{L_{2\mu\mathrm{m}}}{L_{24\mu\mathrm{m}}} \,=\,  4.984^{+1.147}_{-0.836}
    ~,
\end{equation}
\noindent where we report the bootstrapped median value with 16th to 84th percentiles uncertainties. To summarise, \textit{we propose that the near-to-mid infrared luminosity ratio $\mathcal{R}_{\rm IR}$ may help distinguish impostors from true PSBs.}

In Figure~\ref{fig:PSBs_Hdelta}, we compare the Lick H$\delta_{\rm A}$ and D$_{\rm n}4000$ of our sample of PSBs in \firebox (using different selection criteria; fiducial aperture of 7 kpc) to recent observations \citet{French2015,Wild2020,Suess2022}. Overall, \firebox PSBs have values of Lick H$\delta_{\rm A}$ and D$_{\rm n}4000$ that are in reasonable agreement with observations, especially for PSBs selected with the fiducial criterion from \citet{Kriek2010}. The low-SFR PSBs in \firebox, which may be the most representative analogues of observed PSBs, provide the closest match to \squiggle galaxies, with both Lick H$\delta_{\rm A}$ and D$_{\rm n}4000$ agreeing within $\sim 1\sigma$. On average, \firebox PSBs show larger Lick H$\delta_{\rm A}$ and D$_{\rm n}4000$ compared to the bulk of the overall population of \firebox galaxies. The average age of a galaxy's stellar population increases with decreasing Lick H$\delta_{\rm A}$ and increasing D$_{\rm n}4000$. PSBs in \firebox that are selected with the fiducial \citet{Kriek2010} criterion have a lower average Lick H$\delta_{\rm A}$ and D$_{\rm n}4000$ than PSB in the sample of \citet[][]{Suess2022} (selected at a similar redshift and with the same selection criterion), implying that \firebox PSBs have relatively younger stellar populations. This discrepancy could arise from the lower average stellar mass of \firebox galaxies compared to those analysed by \citet{Suess2022}.

In Figure~\ref{fig:PSBs_Halpha}, we compare the distributions of the ratio between the emergent and intrinsic $\rm H\alpha$ luminosities, for PSBs and non-PSBs of similar stellar mass. The ratio $L_{\rm H\alpha}/L_{\rm H\alpha,0}$ provides an estimate of the effective attenuation of the nebular $\rm H\alpha$ emission, with smaller values corresponding to stronger suppression of the emergent line luminosity. Each of the 14 projections available for each galaxy individually contribute to the distributions. Only the projections with SEDs satisfying the \citet{Kriek2010} selection criterion contribute to the PSBs distribution. We find that PSBs have $\rm H\alpha$ emission that is more strongly attenuated, with an average $L_{\rm H\alpha}/L_{\rm H\alpha,0}\simeq 0.1$ compared to the average $L_{\rm H\alpha}/L_{\rm H\alpha,0}\simeq 0.3$ of non-PSBs with similar stellar masses. This result supports the scenario where PSB in \firebox include impostor galaxies in which ongoing star formation is partially obscured. PSBs with genuinely low intrinsic SFRs consistently exhibit $H\alpha$ absorption in both their intrinsic and emergent spectra.

\begin{figure*}
    \centering
    \includegraphics[width=\hsize]{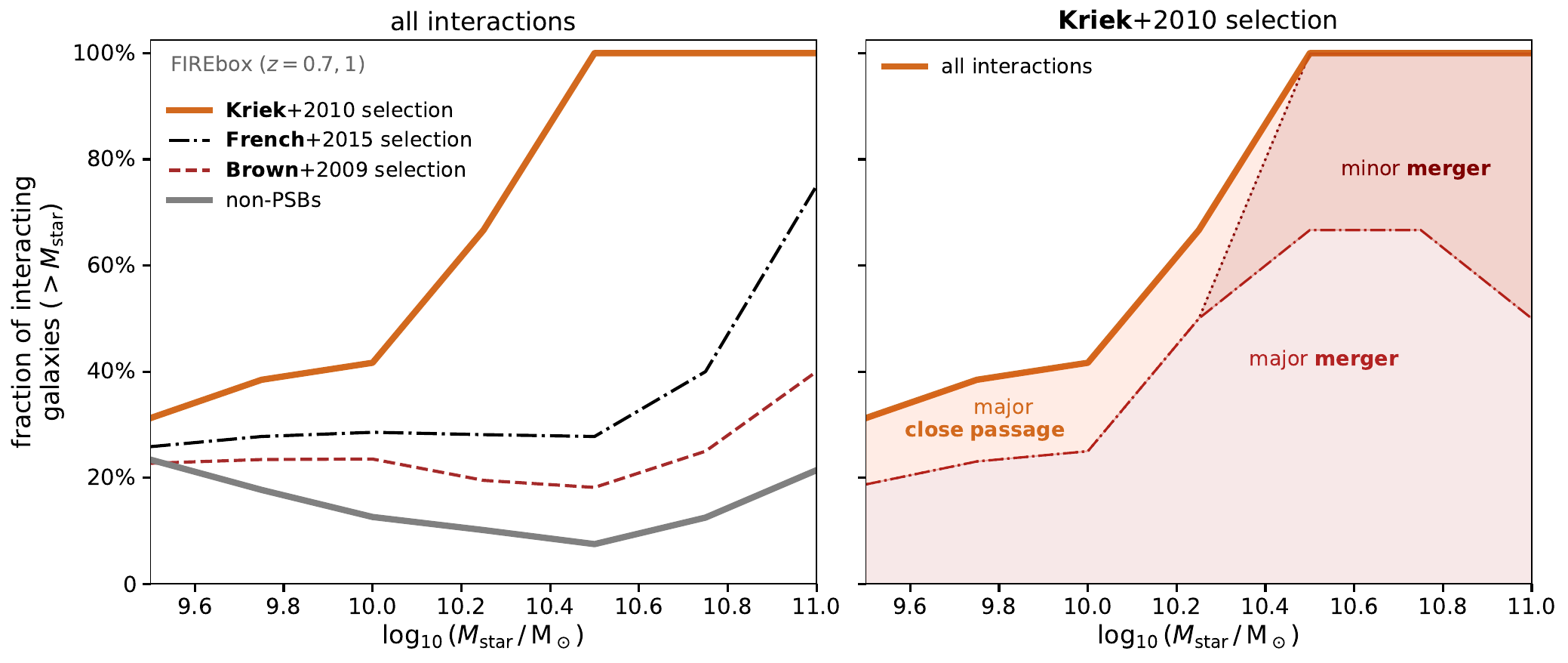}
    \caption{Cumulative fraction of \firebox galaxies that experienced an interaction in the past Gyr, as a function of their stellar mass $M_{\rm star}$. In the left panel, we show the interacting fraction for \firebox PSBs selected according to different criteria and non-PSBs (grey, solid line). For PSBs, we show results relative to the \citet{Kriek2010} selection criterion based on their rest-frame optical photometry (orange, solid line), and the spectroscopy-based criteria following \citet[][]{Brown2009} (red, dashed line) and \citet[][]{French2015} (black, dash-dotted line). The fraction of interacting PSBs is increasingly higher than that of non-PSB with increasing stellar mass. Low-mass ($M_{\rm star}\lesssim 10^{10}\Msun$) PSBs have interaction fractions higher but comparable to those of non-PSBs, whereas high-mass ($M_{\rm star}\gtrsim 10^{10}\Msun$) are significantly more likely interacting than non-PSBs, with an interaction fraction of $100\percent$ at $M_{\rm star}\gtrsim 3\times 10^{10}\Msun$. PSBs selected with spectroscopic criteria have interacting fractions closer to those of non-PSBs compared to PSBs selected based on photometry. In the right panel, we show the kind of interactions of PSBs selected with the fiducial criterion from \citet{Kriek2010}. Major mergers have a dominant role at all masses, with major close passages and minor mergers being of secondary importance.}
    \label{fig:PSBs_merger}
\end{figure*}

\subsection{The role of galaxy interactions}\label{sec:results_int}
In Figure~\ref{fig:PSBs_merger} (left panel), we show the (cumulative) fraction of \firebox galaxies that have experienced a galaxy interaction (see Section~\ref{sec:methods_interactions}) in the past time interval of $\Delta t_{\rm int} = 1\Gyr$. In general, in our sample, \textit{a larger fraction of \firebox PSBs have experienced interactions in the past Gyr compared to non-PSBs}. \textit{This is especially evident for high stellar masses} ($M_{\rm star}\gtrsim 10^{10}\Msun$) and for the fiducial sample of PSBs (i.e. selected with the \citet{Kriek2010} criterion). Low-mass PSBs are temporarily quenched by stellar feedback following an intense starburst event and observed during the periods of low SFR between deep oscillations in their star-formation histories \citep[akin to the ‘breathing' mode described in][]{Cenci2024a}. Strong starburst events are commonly driven by mergers and galaxy interactions in massive systems \citep[$M_{\rm star}\gtrsim 10^{10}\Msun$; see, e.g.,][]{Cenci2024a}, likely leading to a phase with observational features akin to those of PSBs. Furthermore, in Figure~\ref{fig:PSBs_merger} (right panel), we report the dominant interaction class (e.g., major and minor mergers; see Section~\ref{sec:methods_interactions} for definitions) among PSBs in the fiducial sample. \textit{Major mergers play an important role throughout the entire mass range we considered}. The remainder ($\sim 30\percent$) of the interacting PSBs in the fiducial sample mainly undergo minor mergers and major close passages, for $M_{\rm star}\gtrsim 3\times 10^{10}\Msun$ and $M_{\rm star}\lesssim 3\times 10^{10}\Msun$, respectively. These results are consistent with previous numerical work on simulated PSBs, that report that major-mergers (minor/micro-mergers) are important PSB drivers at low (high) redshift \citep[e.g.,][who analysed PSBs in the EAGLE simulation]{Davis2019}.
    
\section{Discussion}\label{sec:discussion}

We have found that a large fraction of photometrically-PSBs selected in \firebox are still actively forming stars. Therefore, in this section, we will adopt a Bayesian framework to to investigate the implications of high impostor fractions. In particular, we estimate the fraction of impostors among selected PSBs and the importance of black hole feedback in originating true PSBs in the Universe. To do so, we will use results from both \firebox and observations. We use \firebox to estimate the fraction of selected PSBs among star-forming galaxies and that of galaxies that are (temporarily) quenched due to stellar feedback (alone). From observations, we primarily consider the fraction of selected PSBs and quenched galaxies.

Let us assume that the entire galaxy population is made up of three disjoint, exhaustive classes of galaxies: star-forming galaxies (SF), true post-starburst galaxies\footnote{In this framework, the population of \textit{true} PSBs (here labelled GV) would not include the impostor PSBs that we observe in \firebox, that are instead contributing to the population of star-forming galaxies (here labelled SF).} (here labelled as ‘green-valley' galaxies; GV), and (long-term) quiescent galaxies (Q). These sub-populations represent a fraction $\mathcal{P}\left(\text{SF}\right)$, $\mathcal{P}\left(\text{GV}\right)$, and $\mathcal{P}\left(\text{Q}\right)$ of the entire population, respectively. Furthermore, let us consider a sub-population of PSBs that are selected among GV and SF galaxies, make up a fraction $\mathcal{P}\left(\text{PSB}\right)$ of all galaxies.

In general, \firebox under-predicts the fraction of quiescent galaxies, likely due to the lack of feedback from active galactic nuclei (AGN) and black hole physics \citep[see][]{Feldmann2023}. Nonetheless, \firebox yields realistic star-forming galaxies, reproducing key properties and scaling relations (e.g., the star-forming main sequence and gas sequence), despite not being calibrated to do so. Therefore, the simulation possibly under-predicts the fraction of PSBs that contribute to the GV population, but we rely on it it to estimate of the fractional contribution of PSBs to the SF population, $\mathcal{P}\left(\text{PSB}\lvert\text{SF}\right)$.

\subsection{Impostor fraction}\label{sec:discussion_impostor}
In this section, we will use our findings with \firebox to estimate the impostor fraction in observed samples of PSBs. Our aim is to provide a framework to infer how many of the selected PSBs are expected to be misclassified (i.e. impostors) in observations once other, independent statistics (e.g., the fraction of quenched galaxies) are given. In the framework introduced above, impostor PSBs are galaxies that are selected as PSBs (e.g., based on their photometry, as in the fiducial PSB sample in \firebox) but that are not GV galaxies and are instead SF galaxies. In our notation, impostor PSBs represent a fraction $\Fim\equiv\mathcal{P}\left(\text{Im}\lvert\text{PSB}\right)$ of selected PSBs, given by the following expression (see Appendix~\ref{app:derive_all} for a derivation):
\begin{equation}
    \Fim
    =\frac{\mathcal{P}\left(\text{PSB}\lvert\text{SF}\right)\left[1-\dfrac{\mathcal{P}\left(\text{PSB}\right)}{\mathcal{P}\left(\text{PSB}\lvert\text{GV}\right)}-\mathcal{P}\left(\text{Q}\right)\right]}{\,\mathcal{P}\left(\text{PSB}\right)\left[1-\dfrac{\mathcal{P}\left(\text{PSB}\lvert\text{SF}\right)}{\mathcal{P}\left(\text{PSB}\lvert\text{GV}\right)}\right]\,}
    ~.\label{eqn:impostor_1}
\end{equation}
\noindent We insert the various terms in the equation above using either data from \firebox or from observations, and estimate the fraction of PSBs selected among SF galaxies with \firebox. We calculate that $2.2\percent$ of star-forming galaxies in the simulation are selected as PSBs, i.e. $\mathcal{P}\left(\text{PSB}\lvert\text{SF}\right) = 0.022$. From observations, we can obtain estimates for the fraction of PSBs selected among all galaxies, i.e. $\mathcal{P}\left(\text{PSB}\right)$, that ranges from about $1\percent$ at $z\simeq 0$ to about $10\percent$ at $z\simeq2$ \citep[e.g.,][]{Wild2016,Rowlands2018a,French2021,Suess2022}. Observations also provide estimates for the fraction $\mathcal{P}\left(\text{Q}\right)$ of quenched galaxies among all galaxies. For stellar masses $M_{\rm star}\simeq 10^{9.7-11}\Msun$, we have $\mathcal{P}\left(\text{Q}\right)\lesssim 0.5$ at $z\lesssim 1$ \citep[e.g.,][]{Wild2016,Leja2022}. In general, the exact values reported for $\mathcal{P}\left(\text{PSB}\right)$ and $\mathcal{P}\left(\text{Q}\right)$ display clear trends with respect to both redshift and stellar mass \citep[e.g.,][]{Wild2016,Rowlands2018a}.Here, we assume that galaxies in the GV population are always observationally identified as PSBs, i.e. $\mathcal{P}\left(\text{PSB}\lvert\text{GV}\right) = 1$. In summary, assuming that among all galaxies we have ($i$) a fraction $\mathcal{P}\left(\text{PSB}\right)=0.05$ being selected as PSBs, and ($ii$) a fraction $\mathcal{P}\left(\text{Q}\right)=0.5$ of quenched galaxies, together with ($iii$) a fraction $\mathcal{P}\left(\text{PSB}\lvert\text{SF}\right) = 0.022$ of SF galaxies being (erroneously) selected as PSBs, we estimate\footnote{The inferred fraction of impostor PSBs crucially depends on the estimate for $\mathcal{P}\left(\text{PSB}\right)$ from observations and on $\mathcal{P}\left(\text{PSB}\lvert\text{SF}\right)$ from \firebox. We note that some sets of values for $\mathcal{P}\left(\text{PSB}\right)$, $\mathcal{P}\left(\text{PSB}\lvert\text{SF}\right)$, and $\mathcal{P}\left(\text{Q}\right)$ would lead to an estimated $\Fim>1$, implying that these very combinations are incompatible. More stringent constraints from future observations on these quantities will considerably improve the estimate of $\Fim$.} a fraction of impostor PSBs $\Fim\gtrsim 0.68$.

\begin{figure*}
    \centering
    \includegraphics[width=\textwidth]{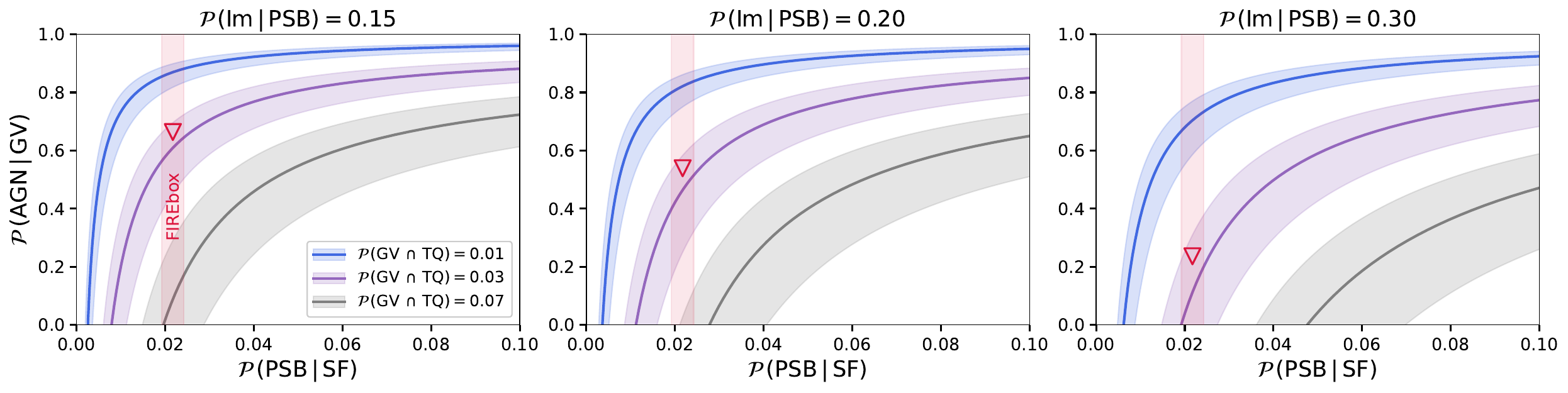}
    \caption{Fraction of true PSBs that are expected to be quenched primarily due to black hole feedback, $\mathcal{P}\left(\text{AGN}\lvert\text{GV}\right)$, as a function of the fraction of PSBs selected among star-forming galaxies, $\mathcal{P}\left(\text{PSB}\lvert\text{SF}\right)$, according to Equation~\eqref{eqn:agn_fraction}. Different lines refer to different values for the fraction of true PSBs that are temporarily quenched due to stellar feedback, $\mathcal{P}\left(\text{GV}\lvert\text{TQ}\right)\mathcal{P}\left(\text{TQ}\right)=\mathcal{P}\left(\text{GV}\,\cap\,\text{TQ}\right)=0.01,0.03,0.07$ (blue, purple, and grey lines, respectively). Different panels assume a different fraction of PSB impostors $\Fim\equiv \mathcal{P}\left(\text{Im}\lvert\text{PSB}\right)=0.15,0.2,0.3$ (from left to right). Results for $\Fim\gtrsim0.4$ are not shown, as we would get $\mathcal{P}\left(\text{AGN}\lvert\text{GV}\right)\simeq 0$. Shaded areas represent the variation in $\mathcal{P}\left(\text{AGN}\lvert\text{GV}\right)$ due to a variation in the assumed fraction of quenched galaxies $\mathcal{P}\left(\text{Q}\right)=0.1-0.5$, where solid lines refer to $\mathcal{P}\left(\text{Q}\right)=0.3$. The red shaded area highlights the value of $\mathcal{P}\left(\text{PSB}\lvert\text{SF}\right)$ estimated with our PSB samples from \firebox ($z=0.7,1$; $M_{\rm star}>3\times 10^{9}\Msun$). Furthermore, we mark with a red triangle the value of $\mathcal{P}\left(\text{AGN}\lvert\text{GV}\right)$ computed by estimating $\mathcal{P}\left(\text{GV}\,\cap\,\text{TQ}\right)$ with \firebox as ${P}\left(\text{q-ASB}\right)=0.026$.}
    \label{fig:PSBs_AGN}
\end{figure*}


We can rearrange Equation~\eqref{eqn:impostor_1} to give an expression for the overall fraction of GV galaxies, $\mathcal{P}\left(\text{GV}\right)$ (see Appendix~\ref{app:derive_all} for a derivation), where we do not need to make assumptions on the value of $\mathcal{P}\left(\text{PSB}\right)$. We have:
\begin{equation}
    \mathcal{P}\left(\text{GV}\right)
    = \dfrac{1-\mathcal{P}\left(\text{Q}\right)}{\,1\,+\,\dfrac{\mathcal{P}\left(\text{PSB}\lvert\text{GV}\right)}{\mathcal{P}\left(\text{PSB}\lvert\text{SF}\right)}}\,\left[\dfrac{1-\Fim}{\Fim}\right]
    ~.\label{eqn:impostor_3}
\end{equation}
\noindent Using infra-red and optical measurement, \citet{Baron2023} show that about $25-30\percent$ of PSBs selected from observations likely host significant levels of obscured, on-going star formation, and that optical studies alone are not sufficient to fully characterise the evolution through the PSB-phase. Based on these results, we can set\footnote{We note that the impostor fraction inferred by \citet{Baron2023} is 
based on spectroscopic selection criteria applied to relatively low-redshift galaxies, and it is thus not treated as a direct calibration of the \firebox impostor population. Instead, we use it only as an observationally motivated reference value in the present Bayesian framework.} $\Fim=0.27$ and $\mathcal{P}\left(\text{Q}\right)=0.5$ \citep[see, e.g.,][]{Wild2016,Leja2022}, and from \firebox, we have again $\mathcal{P}\left(\text{PSB}\lvert\text{SF}\right)=0.022$. Here, we still assume $\mathcal{P}\left(\text{PSB}\lvert\text{GV}\right)=1$. Plugging these estimates into Equation~\eqref{eqn:impostor_3} above yields $\mathcal{P}\left(\text{GV}\right)=0.029$, implying that PSBs selected in observations should represent a fraction $\mathcal{P}\left(\text{PSB}\right)\gtrsim 0.11$ of the entire galaxy population.

Rather than assuming a constant fraction of true GV galaxies selected as PSBs $\mathcal{P}\left(\text{PSB}\lvert\text{GV}\right)=1$, we can estimate it within the simulation as the fraction $\mathcal{P}\left(\text{PSB}\lvert\text{q-ASB}\right) = 0.381$ of selected PSBs among (temporarily) quenched \firebox galaxies that experienced a recent drop in their SFR. In this case, GV galaxies would represent a fraction $\mathcal{P}\left(\text{GV}\right)=0.095$ of the overall galaxy population and we get $\mathcal{P}\left(\text{PSB}\right)\gtrsim 0.10$.

\subsection{The role of black hole feedback}\label{sec:discussion_agn}
\firebox does not include feedback from active galactic nuclei (AGN), nor black hole physics in general, allowing us to explore the effects of other processes in producing the identified population of PSBs in the simulation. In our simulation, stellar feedback is not sufficient to effectively quench galaxies and explain the high fraction of (true) PSBs that is inferred from observations. We suggest that another (non-stellar) feedback channel is the \textit{main} quenching channel for those systems. Most likely, this is due to black hole feedback from active-galactic nuclei (AGN). For instance, if the fraction of impostor PSBs is about $27\percent$, due to obscuration and projection effects (following \citet{Baron2023}; however, see our estimates from the previous Section~\ref{sec:discussion_impostor}), then about $63\percent$ of true PSBs (restricted to central galaxies) in the Universe must be quenched by additional mechanisms other than stellar feedback or gas consumption \citep[e.g.,AGN feedback; e.g.,][]{Hopkins2006a,Alatalo2014,Smercina2018,Zheng2020}. In this section, we aim to assess the fraction of PSBs, selected in observations, where AGN feedback is the main responsible for driving the quenching of these galaxies.

Let us consider the framework introduced in the previous Section~\ref{sec:discussion_impostor} and distinguish two (mutually exclusive, and exhaustive) hypothetical classes of true GV galaxies: those that are temporarily quenched (TQ) due to stellar feedback, and those that are rapidly quenched by AGN feedback. The fractional contribution of these two complementary classes to the overall GV population are represented by $\mathcal{P}\left(\text{TQ}\lvert\text{GV}\right)$ and $\mathcal{P}\left(\text{AGN}\lvert\text{GV}\right)$, respectively, such that:
\begin{equation}
    \mathcal{P}\left(\text{TQ}\lvert\text{GV}\right) + \mathcal{P}\left(\text{AGN}\lvert\text{GV}\right) = 1
    ~.
\end{equation}
\noindent As a caveat to this approach, both feedback channels can in principle be in place at the same time and have a complex interplay. Our classification is solely based on what process would theoretically be the dominant one in affecting the galaxy evolution.

Assume now that the fraction of AGN-quenched galaxies is vanishingly small among star-forming galaxies (i.e. $\mathcal{P}\left(\text{AGN}\lvert\text{SF}\right) = 0$) and maximal for quenched galaxies (i.e. $\mathcal{P}\left(\text{AGN}\lvert\text{Q}\right) = 1$). From Bayes' theorem, we have:
\begin{equation}
    \mathcal{P}\left(\text{AGN}\lvert\text{GV}\right)=1-\frac{\mathcal{P}\left(\text{GV}\cap\text{TQ}\right)}{\mathcal{P}\left(\text{GV}\right)}
    ~,\label{eqn:agn_fraction}
\end{equation}
\noindent where $\mathcal{P}\left(\text{GV}\right)$ is given by Equation~\eqref{eqn:impostor_3} in the previous section. From observational constraints, we now assume $\Fim=0.27$ \citep[][]{Baron2023} and $\mathcal{P}\left(\text{Q}\right)=0.5$ \citep[see, e.g.,][]{Wild2016,Leja2022}, whereas from \firebox we can estimate $\mathcal{P}\left(\text{PSB}\lvert\text{SF}\right)=0.022$ and $\mathcal{P}\left(\text{PSB}\lvert\text{GV}\right) \simeq \mathcal{P}\left(\text{PSB}\lvert\text{q-ASB}\right) = 0.29$. Furthermore, the fraction of galaxies that is temporarily quenched due to stellar feedback, among all galaxies at $z\sim 0.5-1$, can be estimated with \firebox as the total fraction of \text{q-ASBs}: $\mathcal{P}\left(\text{GV}\cap\text{TQ}\right)\sim\mathcal{P}\left(\text{q-ASB}\right)\simeq 0.026$. Plugging these numbers into Equation~\eqref{eqn:agn_fraction}, we obtain that \textit{about $33.6\percent$ of real PSBs must be quenched due to AGN feedback}. The variation of this estimate are illustrated in Figure~\ref{fig:PSBs_AGN}, where we show the dependence of $\mathcal{P}\left(\text{AGN}\lvert\text{GV}\right)$ on the other relevant quantities in Equation~\eqref{eqn:agn_fraction}. The number of PSB candidates in \firebox is too small to derive a robust mass-binned estimate for all the relevant terms in Equation~\eqref{eqn:agn_fraction}. We therefore report an integrated value for $\mathcal{P}\left(\text{PSB}\lvert\text{SF}\right)$, for which we find no clear statistically significant stellar-mass dependence in the current sample.

Recent observational work by \citet{Zhu2025} that massive PSBs in the \squiggle survey have consistently weak $\mathrm{H}\alpha$ emission, suggesting that their ongoing star-formation activities are genuinely low despite the presence of substantial molecular gas reservoirs. Diagnostic line ratios further suggest that even the small amount of $\mathrm{H}\alpha$ emission detected in \squiggle PSBs is often powered by non–star-forming ionization sources (e.g., AGN), implying that $\mathrm{H}\alpha$-based SFRs may represent upper limits in their quenched systems. Furthermore, $\mathrm{H}\alpha$ always largely over-predicts the SFR of their PSBs compared to the estimates given based on the SEDs of the same objects \citep[][]{Bezanson2022}. However, the latter could be under-estimated, implying that many \squiggle PSBs are more actively star-forming than previously thought \citep[see][]{Setton2025}. The results of \citet{Zhu2025} prove that obscured star formation may not be the main contribution to the occurrence of impostors among selected PSBs. In this context, our results strongly suggest that AGN feedback is required in order to reproduce the observed population of PSBs.

\section{Summary and Conclusions}\label{sec:conclusions}
We studied the nature of post-starburst galaxies (PSBs) in the \firebox cosmological simulation, identified based on, primarily, their photometric properties following observational criteria \citep[e.g.,][]{Kriek2010}. We considered all central galaxies at $z=0.7$ and $z=1$ with stellar masses $M_{\rm star}>3\times 10^9\Msun$ in \firebox, for a total of 310 galaxies, and modelled spectral-energy distributions (SEDs) and photometry using the radiative transfer simulation code \skirt \citep[][]{Camps_and_Baes2020}. Mock SEDs were generated for 14 viewing angles per galaxy, allowing us to quantify the impact of projection effects and dust geometry on PSB selection. We used predictions from \firebox to constrain the fraction of selected PSBs that may be impostors, and to assess which quenching channels are required to reproduce a realistic PSB population. Our main findings and conclusions are summarised as follows:
\renewcommand{\labelitemi}{$\bullet$}
\begin{itemize}

    \item[($i$)]
        PSBs selected with the fiducial photometric criterion represent about $3.1\percent$ of all galaxies with $M_{\rm star}>3\times 10^9\Msun$ $z=0.7$ and $z=1$ in \firebox. 
        Restricting the parent sample to star-forming galaxies, this fraction becomes $2.2\percent$, implying that a large fraction of PSBs in \firebox are still actively star-forming. The overall PSB fraction depends on the aperture used to measure the mock photometry, varying from about $2$ to $4\percent$ for the \citet{Kriek2010} colour selection when the aperture radius is varied from $12$ to $1\kpc$ (see Figure~\ref{fig:PSBs_fraction}). Spectroscopic selection criteria generally yield somewhat larger PSB fractions, up to about $7.5\percent$ for apertures larger than about $6\kpc$.

    \item[($ii$)]
        A large fraction of photometrically PSBs selected in \firebox are not genuinely quenched systems. About $77.6\percent$ of selected PSBs have experienced a recent significant drop in their SFR, but only $31.8\percent$ are temporarily quenched (\text{q-ASBs}), with $\mathrm{sSFR}<3\times 10^{-11}~\mathrm{yr}^{-1}$ for $\tavg=20\Myr$. About $2.2\percent$ of all star-forming galaxies in our sample have optical photometric properties consistent with PSBs. Furthermore, about $68.2\percent$ of selected PSBs are either still actively star-forming or have not experienced a sufficiently strong recent decline in SFR. The latter estimate is likely an upper limit due to the lack of massive black hole physics in the simulation (see point ($x$) below).

    \item[($iii$)]
        Many PSBs in \firebox lie within the characteristic scatter of the star-forming main sequence, for both SFR averaging times of $\tavg=20\Myr$ and $\tavg=100\Myr$ (see Figure~\ref{fig:PSBs_SFMS}). The PSBs with $\mathrm{sSFR}<3\times 10^{-11}~\mathrm{yr}^{-1}$ (for $\tavg=20\Myr$) are preferentially systems that have experienced a rapid decline in their recent star formation, typically within the past $\lesssim 100\Myr$. Therefore, these low-SFR objects represent the closest analogues to true PSBs in \firebox, although their quenching is generally temporary.

    \item[($iv$)]
        PSBs in \firebox commonly retain substantial gas reservoirs. Their total gas and molecular gas masses are often comparable to those of star-forming galaxies with similar stellar masses (see Figure~\ref{fig:PSBs_H2}). Star-forming PSBs (impostors) have molecular-gas depletion times of about $400\Myr$, similar to normal star-forming galaxies in \firebox, whereas low-SFR PSBs have longer depletion times, of order $\sim 2\Gyr$ (see Figure~\ref{fig:PSBs_tdepl}). PSBs selected in \firebox have molecular gas fractions and recent formation timescales broadly consistent with galaxies in the \squiggle sample of PSB candidates at $z\simeq0.7$ with confirmed molecular gas detections (see Figure~\ref{fig:PSBs_H2_vs_ages}). However, gas-rich PSBs in \firebox are almost exclusively impostors with ongoing star formation. Conversely, low-SFR PSBs in \firebox tend to have molecular gas masses below typical detection limits. This suggests that molecular gas-rich observed PSBs may include a substantial contribution from impostors, whereas truly quenched PSBs may be preferentially gas-poor or difficult to detect.

    \item[($v$)]
        The star-formation histories of PSBs in \firebox show that many of them undergo bursty cycles, with phases of enhanced star formation followed by rapid drops in both SFR and molecular gas fraction (see Figure~\ref{fig:PSBs_SFHs}). In low-mass galaxies, these cycles are consistent with temporary quenching driven by stellar feedback and gas consumption, similar to the breathing-galaxy behaviour discussed in \citet{Cenci2024a}. In more massive systems, recent bursty behaviour is more often associated with galaxy interactions.

    \item[($vi$)]
        Mid-infrared emission provides a useful diagnostic for distinguishing impostor PSBs from genuinely low-SFR PSBs. Impostor PSBs have MIR luminosities similar to those of non-PSB star-forming galaxies with comparable NIR luminosities. By contrast, low-SFR PSBs have weaker $24~\mu$m emission at fixed $2~\mu$m luminosity (see Figure~\ref{fig:PSBs_IR}). We find that a rest-frame near-to-mid infrared luminosity ratio of $\mathcal{R}_{\rm IR}\simeq5.07$ provides an approximate separation between low-SFR PSBs and star-forming impostors in \firebox.

    \item[($vii$)]
        PSBs in \firebox experience a stronger average attenuation of $\rm H\alpha$ emission, supports the interpretation that some \firebox PSBs are impostors with ongoing but partially obscured star formation (see Figure~\ref{fig:PSBs_Halpha}). In contrast, genuinely low-SFR PSBs exhibit $\rm H\alpha$ absorption in both their intrinsic and emergent spectra.

    \item[($viii$)]
        PSBs in \firebox are more likely than non-PSBs to have experienced galaxy interactions in the past Gyr, especially at $M_{\rm star}\gtrsim10^{10}\Msun$ (see Figure~\ref{fig:PSBs_merger}). At the highest stellar masses considered, the interaction fraction of PSBs approaches unity. Major mergers provide the dominant contribution to the interaction history of PSBs across the explored mass range, while minor mergers and major close passages play a secondary role.

    \item[($ix$)]
        Using a Bayesian framework, we estimated the impostor fraction expected in observed PSB samples. Combining the fraction of star-forming galaxies selected as PSBs in \firebox, $\mathcal{P}\left(\text{PSB}\lvert\text{SF}\right)=0.022$, with observationally motivated values for the total PSB and quenched fractions, we infer that the impostor fraction in observations may be large, potentially $\Fim\gtrsim0.68$ for representative choices of $\mathcal{P}\left(\text{PSB}\right)=0.05$ and $\mathcal{P}\left(\text{Q}\right)=0.5$. This estimate is sensitive to the assumed PSB fraction, quenched fraction, and selection efficiency of true PSBs.

    \item[($x$)]
        Since \firebox does not include black hole physics, it provides a useful reference for the PSB population that can be produced by stellar feedback, gas consumption, and galaxy interactions alone. We find that these processes can produce galaxies with PSB-like photometric properties, but they do not naturally generate a sufficiently large population of long-lived, truly quenched PSBs. Within our Bayesian framework, and adopting observationally motivated impostor fractions from \citet{Baron2023}, we estimate that about $33.6\percent$ of true PSBs in the Universe may require an additional quenching mechanism beyond stellar feedback, most plausibly massive black hole feedback (see Figure~\ref{fig:PSBs_AGN}).

\end{itemize}

In summary, we show that \textit{galaxies with significant levels of star formation can masquerade as PSB impostors}, displaying PSB-like SEDs and optical colours while having SFRs, molecular gas reservoirs, and depletion times typical of star-forming galaxies. Our results suggest that the fraction of impostors is likely substantial, and we propose that the near-to-mid infrared luminosity ratio may help distinguish them from true PSBs. Furthermore, stellar feedback and gas consumption can temporarily suppress star formation and generate PSB-like systems, especially in lower-mass galaxies, while galaxy interactions and major mergers become increasingly important at higher stellar masses. However, these channels alone are unlikely to explain the full population of genuinely quenched PSBs inferred from observations. If observed PSB samples contain only a moderate impostor fraction, then an additional quenching mechanism, most likely black hole feedback, is required to reproduce a realistic PSB population.

Future work using larger simulated volumes, wider ranges in stellar mass and redshift, and simulations including explicit black hole growth and feedback processes will be essential to determine the physical origin of PSBs. On the observational side, multi-wavelength diagnostics combining optical spectra, infrared emission, radio continuum, and molecular gas measurements will be crucial for separating true PSBs from star-forming impostors.

\section*{Acknowledgements}
We thank the anonymous referee for their constructive comments, which greatly improved the present manuscript. EC thanks Papa IV for His countless blessings. RF, LB acknowledge financial support from the Swiss National Science Foundation (grant no PP00P2$\_$194814). EC, RF, MB acknowledge financial support from the Swiss National Science Foundation (grant no 200021$\_$188552). SW received support from NASA grant 80NSSC24K0838. JG gratefully acknowledges financial support from the Swiss National Science Foundation (grant no CRSII5$\_$193826) and funding via STFC grant ST/Y001133/1. DJS is supported by The Brinson Foundation through a Brinson Prize Fellowship grant. LT is grateful to the University of Zurich for supporting this work. We acknowledge PRACE for awarding us access to MareNostrum at the Barcelona Supercomputing centre (BSC), Spain. This research was partly carried out via the Frontera computing project at the Texas Advanced Computing centre. Frontera is made possible by National Science Foundation award OAC-1818253. This work was supported in part by a grant from the Swiss National Supercomputing Centre (CSCS) under project IDs s697 and s698. We acknowledge access to Piz Daint at the Swiss National Supercomputing Centre, Switzerland, under the University of Zurich’s share with the project ID uzh18. This work made use of infrastructure services provided by S3IT (\url{www.s3it.uzh.ch}), the Service and Support for Science IT team at the University of Zurich. All plots were created with the \textsc{matplotlib} library for visualisation with Python \citep{Hunter2007}. This project is part of the FIRE simulation collaboration.

\section*{Data Availability}
The data supporting the plots within this article are available on reasonable request to the corresponding author. A public version of the \textsc{gizmo} code is available at \url{http://www.tapir.caltech.edu/~phopkins/Site/GIZMO.html}.



\bibliographystyle{mnras}
\bibliography{main}


\appendix
\section{Derivation of the fraction of impostor PSBs}\label{app:derive_all}
In this appendix, we derive the relevant expressions used in Section~\ref{sec:discussion_impostor}. We assume that the galaxy population is made up of three disjoint, exhaustive classes of galaxies: star-forming galaxies
(SF), ’true’ post-starburst galaxies (labelled as green-valley
galaxies; GV), and long-term quiescent galaxies (Q):
\begin{equation}
    \mathcal{P}\left(\text{SF}\right) + \mathcal{P}\left(\text{GV}\right) + \mathcal{P}\left(\text{Q}\right) = 1
    ~.\label{eqn:app_1}
\end{equation}
\noindent Furthermore, we consider the population of selected PSBs, among star-forming and green-valley galaxies:
\begin{equation}
    \mathcal{P}\left(\text{PSB}\right) = \mathcal{P}\left(\text{PSB}\lvert\text{GV}\right)\mathcal{P}\left(\text{GV}\right) + \mathcal{P}\left(\text{PSB}\lvert\text{SF}\right)\mathcal{P}\left(\text{SF}\right)
    ~.\label{eqn:app_2}
\end{equation}
\noindent In the right-hand side of the equation above, the first term refers to the identified \true PSBs, whereas the second term refers to star-forming galaxies that have been (erroneously) selected as PSBs (impostor PSBs; Im). In this framework, impostor PSBs are those galaxies that are selected as PSBs among star-forming galaxies, and thus are not ’true’ PSBs:
\begin{equation}
    \mathcal{P}\left(\text{Im}\right) = \mathcal{P}\left(\text{PSB}\lvert\text{SF}\right)\mathcal{P}\left(\text{SF}\right)
    ~.\label{eqn:app_3}
\end{equation}
\noindent By definition, we have that $\mathcal{P}\left(\text{PSB}\lvert\text{Im}\right)=1$. Therefore, the fraction of impostors PSBs among PSBs, $\Fim$, is given by:
\begin{equation}
    \Fim \equiv \mathcal{P}\left(\text{Im}\lvert\text{PSB}\right) = \frac{\mathcal{P}\left(\text{Im}\right)}{\mathcal{P}\left(\text{PSB}\right)}
    ~.\label{eqn:app_4}
\end{equation}
\noindent From Equation~\eqref{eqn:app_2}, the fraction of ’true’ PSBs is given by:
\begin{equation}
    \mathcal{P}\left(\text{GV}\right) = \frac{\mathcal{P}\left(\text{PSB}\right) - \mathcal{P}\left(\text{PSB}\lvert\text{SF}\right)\mathcal{P}\left(\text{SF}\right)}{\mathcal{P}\left(\text{PSB}\lvert\text{GV}\right)}
    ~.\label{eqn:app_5}
\end{equation}
\noindent Using Equation~\eqref{eqn:app_1} to replace $\mathcal{P}\left(\text{SF}\right)$, and rearranging, we obtain:
\begin{equation}
    \mathcal{P}\left(\text{GV}\right) = \frac{\mathcal{P}\left(\text{PSB}\right) - \mathcal{P}\left(\text{PSB}\lvert\text{SF}\right)\left[1-\mathcal{P}\left(\text{Q}\right)\right]}{\mathcal{P}\left(\text{PSB}\lvert\text{GV}\right)\left[1-\dfrac{\mathcal{P}\left(\text{PSB}\lvert\text{SF}\right)}{\mathcal{P}\left(\text{PSB}\lvert\text{GV}\right)}\right]}
    ~.\label{eqn:app_6}
\end{equation}
\noindent Plugging Equation~\eqref{eqn:app_6} into Equation~\eqref{eqn:app_5}, and using Equations~\eqref{eqn:app_3} and~\eqref{eqn:app_4}, we obtain the expression for the impostor fraction:
\begin{equation}
    \Fim
    =\frac{\mathcal{P}\left(\text{PSB}\lvert\text{SF}\right)\left[1-\dfrac{\mathcal{P}\left(\text{PSB}\right)}{\mathcal{P}\left(\text{PSB}\lvert\text{GV}\right)}-\mathcal{P}\left(\text{Q}\right)\right]}{\,\mathcal{P}\left(\text{PSB}\right)\left[1-\dfrac{\mathcal{P}\left(\text{PSB}\lvert\text{SF}\right)}{\mathcal{P}\left(\text{PSB}\lvert\text{GV}\right)}\right]\,}
    ~.
\end{equation}

\bsp	
\label{lastpage}
\end{document}